\documentclass[preprint2]{aastex}

\slugcomment{To appear in Astrophysical Journal Letters}

\shorttitle{The star/brown dwarf ratio revisited}
\shortauthors{Scholz et al.}

\begin{document}
\bibliographystyle{apj}


\title{Substellar Objects in Nearby Young Clusters VII: \\The substellar mass function revisited}

\author{Alexander Scholz\altaffilmark{1,2}, Vincent Geers\altaffilmark{1}, 
Paul Clark\altaffilmark{3}, Ray Jayawardhana\altaffilmark{4,**}, Koraljka Muzic\altaffilmark{5}}

\email{aleks@cp.dias.ie}

\altaffiltext{1}{School of Cosmic Physics, Dublin Institute for Advanced Studies, 31 Fitzwilliam Place,
Dublin 2, Ireland}
\altaffiltext{2}{School of Physics and Astronomy, University of St Andrews, North Haugh, St Andrews, Fife KY16 9SS, United Kingdom}
\altaffiltext{3}{Zentrum f{\"u}r Astronomie der Universit{\"a}t Heidelberg, Institut f{\"u}r Theoretische Astrophysik,  
Albert-Ueberle-Str. 2, D-69120 Heidelberg, Germany}
\altaffiltext{4}{Department of Astronomy \& Astrophysics, University of Toronto, 50 St. George Street, Toronto, 
ON M5S 3H4, Canada}
\altaffiltext{5}{European Southern Observatory, Alonso de C\'{o}rdova 3107, Casilla 19, 19001, Santiago, Chile}
\altaffiltext{**}{Principal Investigator of SONYC}

\begin{abstract}
The abundance of brown dwarfs (BDs) in young clusters is a diagnostic of star formation theory. Here we revisit the 
issue of determining the substellar initial mass function (IMF), based on a comparison between NGC1333 and IC348, two 
clusters in the Perseus star-forming region. We derive their mass distributions for a range of model isochrones, varying 
distances, extinction laws and ages, with comprehensive assessments of the uncertainties. We find that the choice of isochrone 
and other parameters have significant effects on the results, thus we caution against comparing IMFs
obtained using different approaches. For NGC1333, we find that the star/BD ratio $R$ is between 1.9 and 2.4, for 
all plausible scenarios, consistent with our previous work. For IC348, $R$ is found to be between 2.9 and 4.0, 
suggesting that previous studies have overestimated this value. Thus, the star forming process generates about 
2.5-5 substellar objects per 10 stars. The derived star/BD ratios correspond to a slope of the power-law mass 
function of $\alpha = 0.7-1.0$ for the 0.03-1.0$\,M_{\odot}$  mass range. The median mass in these clusters -- 
the typical stellar mass -- is between 0.13-0.30$\,M_{\odot}$. Assuming that NGC1333 is at a shorter distance 
than IC348, we find a significant difference in the cumulative distribution of masses between the two clusters, 
resulting from an overabundance of very low mass objects in NGC1333. Gaia astrometry will constrain the cluster 
distances better and will lead to a more definitive conclusion. Furthermore, the star/BD ratio is somewhat 
larger in IC348 compared with NGC1333, although this difference is still within the margins of error. Our 
results indicate that environments with higher object density may produce a larger fraction of very low mass 
objects, in line with predictions for brown dwarf formation through gravitational fragmentation of filaments 
falling into a cluster potential.
\end{abstract}


\section{Introduction}
\label{s1}

Brown dwarfs (BDs) are an ubiquitous outcome of the star formation process. All young
regions investigated so far with sufficient depth host a population of BDs
with masses down to 0.01$\,M_{\odot}$ or even below. The mechanism that governs their
formation, however, remains unknown. It is clear that additional physics needs to be
included in the models for the cloud fragmentation and subsequent evolution, to allow
for the formation of a sizable number of BDs \citep{2007prpl.conf..149B}. Plausible options 
for these processes include fragmentation driven by turbulence, dynamical ejection of embryonic 
BDs from multiple systems, fragmentation of filaments falling into a cluster potential,
or fragmentation of protoplanetary disks, again combined with ejection 
\citep{2007prpl.conf..459W,2008MNRAS.389.1556B,2009MNRAS.392..413S}. 
Young brown dwarfs are a critical population to test the relevance of these processes.

The standard diagnostic to distinguish between theoretical scenarios is the distribution
of stellar and substellar masses after star formation is finished, or the initial mass function (IMF). 
In the literature several parameterisations for the IMF are used, for example a series of power laws 
\citep{2001MNRAS.322..231K} or a lognormal form \citep{2003PASP..115..763C}. For our goal of 
determining the abundance of brown dwarfs, an often used parameterisation of the IMF
is the star/BD ratio $R$, the ratio of the number of objects in the two mass bins from 0.08 to 
1.0$\,M_{\odot}$ and from 0.03 to 0.08$\,M_{\odot}$, where the low mass cutoff as 0.03$\,M_{\odot}$ 
is chosen to assure completeness. The upper mass limit for the stars of 1.0$\,M_{\odot}$ is
to some extent an arbitrary definition, but the relatively small number of higher-mass stars in 
the nearby star forming region assures that this particular choice does not affect the result
much. The star/BD ratio as a metric has the advantage of maximising the sample size in the 
substellar regime and thus minimizing the statistical errors. Because the stellar side of the IMF is 
well-determined for the nearby star forming regions and shows, in the overwhelming majority of regions, 
no evidence for environmental differences \citep{2010ARA&A..48..339B}, any variation in the star/BD ratio 
from one region to another would indicate a change in the BD abundance.

Measuring the substellar mass function and the star/BD ratio is a challenging task.
It needs a consistent survey procedure and a careful analysis of possible 
incompleteness, but the core problem is to estimate the masses. This requires one to make 
assumptions about the distance to and age of the region, as well as the extinction law 
used to deredden the photometry or spectra. Furthermore, the conversion from observed 
quantities to masses can only be done in the framework of a given theoretical isochrone 
and thus depends on the status of the evolutionary models for young stars and brown 
dwarfs.

In previous work within the project SONYC ('Substellar Objects in Nearby Young Clusters')
we have presented tentative evidence for regional differences in the star/BD ratio.
In particular, for the young cluster NGC1333 in the Perseus star forming complex we 
find $R \sim 2-3$ based on a very deep survey with comprehensive spectroscopic follow-up (\citet{2009ApJ...702..805S},
hereafter SONYC-I; \citet{2012ApJ...744....6S}, hereafter SONYC-IV). For other regions we and other groups 
have published 
R-values ranging from 2 to 8 (see SONYC-IV). One of the most extreme cases in the literature 
is the cluster IC348, the second embedded cluster in Perseus and a slightly older sibling to NGC1333 
\citep{2008hsf1.book..308B}, with a star/BD ratio of $R \sim 8$ \citep{2003ApJ...593.1093L,2008ApJ...683L.183A}. 
At face value, this indicates changes in the BD abundance by a factor of 4 occuring within the same
star forming association. So far, however, the uncertainties for the star/BD ratio have not been assessed 
accurately. The goal of this paper is to carry out a benchmark test for the two extreme cases NGC1333 
and IC348 to verify their discrepant star/BD ratios.

\section{The approach}

The core idea of this paper is to determine the mass distribution and the star/BD ratio for two 
young clusters in Perseus and to assess the associated uncertainties. We will start with consistently 
selected samples for the two clusters, i.e. samples which have been put together in a
homogeneous way, to minimize the influence of selection biases and incompleteness. For these samples,
we will then define a consistent way of estimating object masses by comparing photometry with a
given model isochrone. In addition to the choice of the isochrone, the distance to the cluster,
the extinction law, and the age of the region enter as free parameters into this procedure. 
We will define a set of scenarios with plausible choices for these parameters and different isochrones.
We will then estimate object masses and calculate the star/BD ratio and other indicators of the mass
distribution for each of these scenarios. This will yield a useful dataset to discuss the uncertainties 
in these indicators and assess whether there is evidence for regional differences in the substellar
IMF between NGC1333 and IC348.

\subsection{The samples}
\label{s21}

We use the Spitzer-selected sample of young stellar/substellar objects presented by \citet{2009ApJS..184...18G}
which includes our two target regions. For each of the regions, a region of $25'\times 25'$ centered
on the core of the cluster was observed.
The selection in \citet{2009ApJS..184...18G} is based on colours and magnitudes in Spitzer and
2MASS bands from 1 to 24$\,\mu m$. Their multi-colour selection process uses a series of
criteria designed to exclude background stars, non-stellar emission features and 
extragalactic objects. According to their analysis, this process yields only minimal
contamination (a few percent). An earlier version of this process was used in \citet{2008ApJ...674..336G}
in a survey of NGC1333. Assuming that the distances to the clusters are similar, the identical
selection procedure also ensures that the depth and completeness in terms of magnitudes in 
the two samples is comparable. This removes a major obstacle for an accurate comparison of the 
BD abundance.

Since the primary selection criterion is excess emission in the infrared, this sample only contains 
objects with emission from circumstellar material, i.e. either disks or envelopes. It does not include 
disk-less young stars and brown dwarfs (Class III objects). Therefore, to infer star/BD ratios, we have 
to make the assumption that the fraction of objects with disks does not change with object mass 
in the low-mass regime. As recently shown in \citet{2013MNRAS.429..903D}, this is a plausible assumption 
for many star forming regions, including IC348. For NGC1333, there might be a slight mass 
dependence, as the disk fraction in the total Spitzer-selected sample is found to be $83\pm 11$\% 
\citep{2008ApJ...674..336G}, whereas the value for the very low mass objects is only 55-66\% 
(SONYC-IV). This indicates that the star/BD ratio in this cluster could be slightly overestimated. 

The entire Perseus cloud, including the two target clusters have also been observed as part
of the Spitzer Legacy program 'From Cores to Disks' (C2D, PI: N. Evans). Their Perseus YSO catalogue derived
from IRAC photometry is discussed in detail in \citet{2006ApJ...645.1246J}. We prefer to use the
\citet{2009ApJS..184...18G} selection, because it is slightly deeper, which is beneficial for our
purposes. The downside of the 
\citet{2009ApJS..184...18G} sample is the limited spatial coverage. Here the C2D catalogue is useful 
to check the spatial completeness of our samples.

\begin{figure*}
\center
\includegraphics[width=5.5cm,angle=-90]{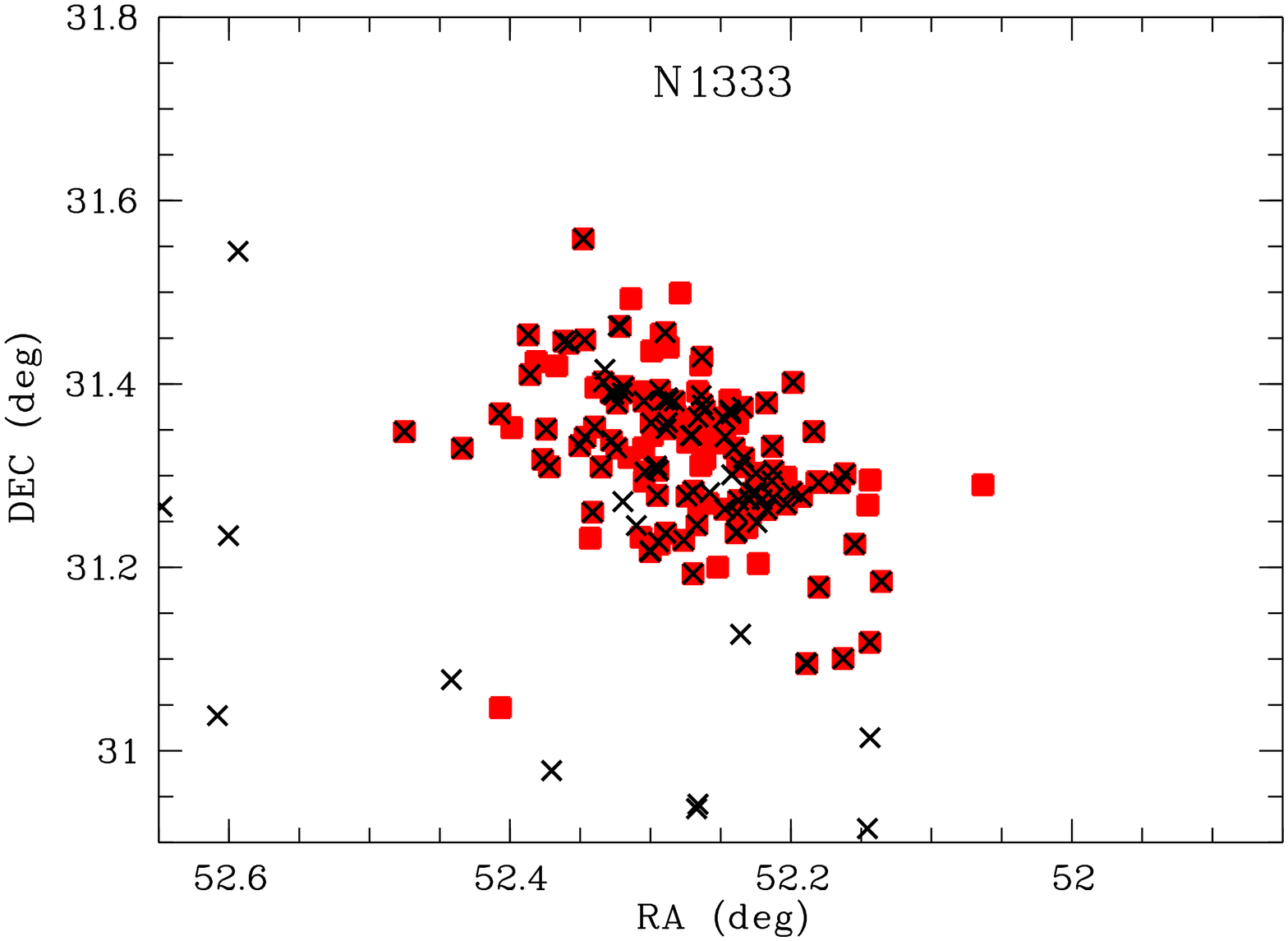} \hfill
\includegraphics[width=5.5cm,angle=-90]{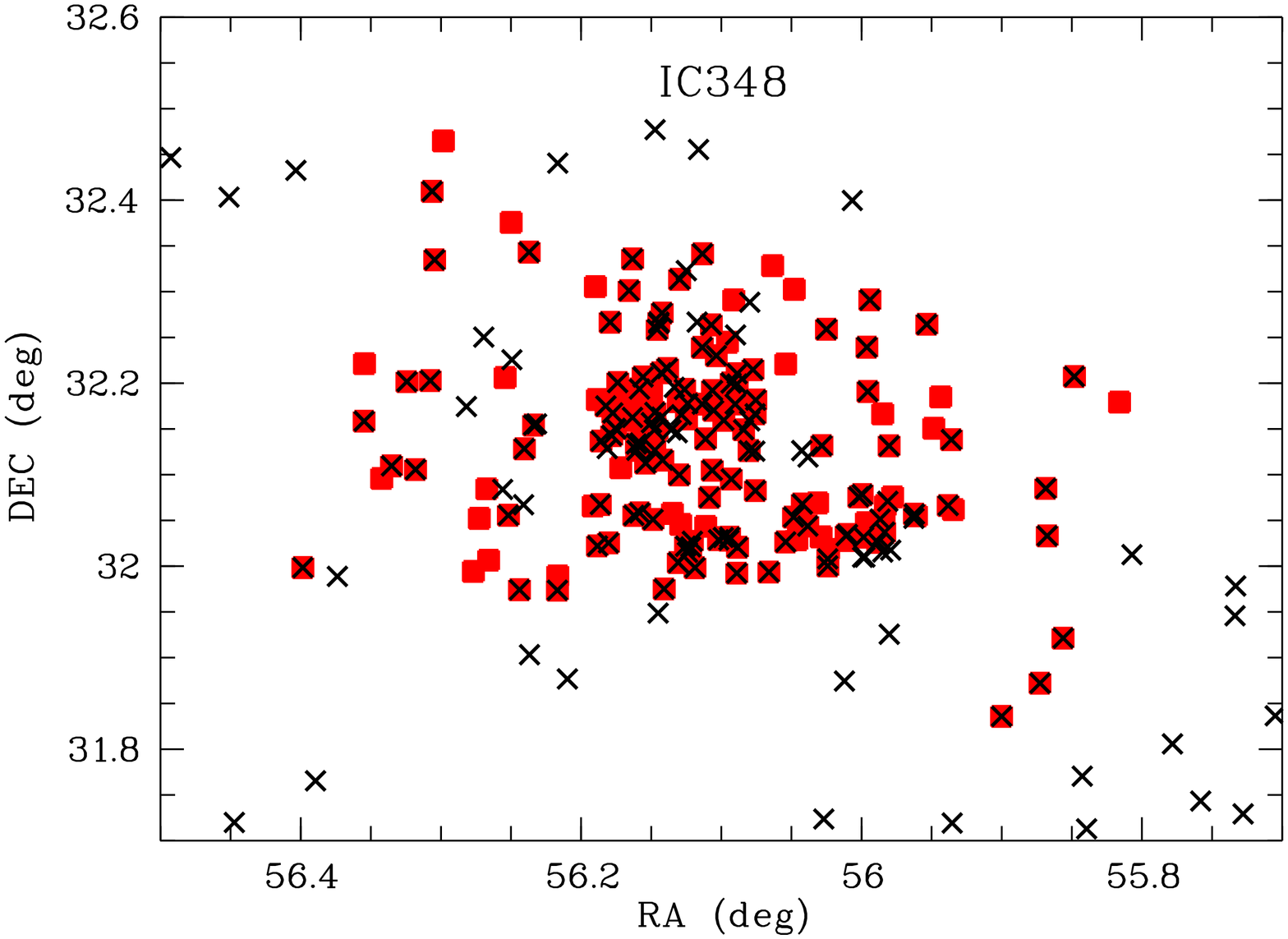} \\
\caption{Spatial distribution of YSOs in the two target regions. Plotted are all objects from the 
\citet{2009ApJS..184...18G} catalogue with red squares For comparison, we also show the YSO candidate
from \citet{2006ApJ...645.1246J} with black crosses. Note that the scale in the two panels is identical.
\label{f0}}
\end{figure*}

In Fig. \ref{f0} we show the spatial distribution of the selected objects in NGC1333 and IC348. With red squares 
we plot the samples from \citet{2009ApJS..184...18G}, with black crosses the C2D sample that covers
the entire Perseus region. The figure shows that the number of YSO candidates outside the region
covered by \citet{2009ApJS..184...18G} is small compared with the total population. This is particularly
true for NGC1333 which shows a very compact profile and can be considered to be spatially complete (see
also SONYC-IV).
For IC348, there is a tail of a YSO population towards the south-west, which represents the transition
region to another densely populated area in the Perseus star forming region \citep{2006A&A...445..999C}.
In addition, there are about 10 objects outside the coverage of the \citet{2009ApJS..184...18G} survey. 
However, there is also a large number of additional YSO candidates only contained in the C2D sample within 
the cluster core. We checked the objects only in C2D and found that they do not show an obvious magnitude 
or extinction bias with respect to the sample we are using, thus, even in case we are missing  
members in the outskirts of the cluster, this is not going to affect our analysis in any significant way.
While \citet{2003AJ....125.2029M} do find a difference in the IMF between the core and the 
halo in IC348, both regions (in their definitions) are within the survey area of \citet{2009ApJS..184...18G}.
We conclude that the samples we are using are not affected by a spatial bias.

Fig. \ref{f0} also illustrates one major difference between IC348 and NGC1333. The core of IC348 has about
twice the diameter of the core of NGC1333 \citep[2.1 vs. 1.2\,pc,][]{2009ApJS..184...18G}, i.e. the cluster
volume in IC348 is about 8 times larger than in NGC1333. On the other hand, IC348 has only about
20\% more YSOs than NGC1333, according to the Spitzer surveys \citep{2008ApJ...683..822J,2009ApJS..184...18G}.
The fraction of diskless Class III objects is higher in IC348, taken that into account the total the 
YSO population in IC348 could be up to twice as large as in NGC1333. This still implies that the
object density in NGC1333 is 4-7 times higher than in NGC1333. Given the age, size, and number of
members in these clusters, this difference is likely to be primordial, and not caused by dynamical
evolution \citep[see Fig. 1 in][]{2012MNRAS.426L..11G}. Hence, these two clusters constitute 
an excellent test case to probe the effects of dynamical interactions and cluster potential on the 
formation of BDs.

\subsection{Estimating masses}
\label{s22}

For the overwhelming majority of young objects, masses can only be estimated indirectly by comparing an 
observed quantity with predictions from theoretical isochrones for a given age. For the observed 
quantity, there are two options, either the effective temperature or the luminosity (or a photometric 
proxy). The luminosity has the problem that model derivations are sensitive to the age for pre-main 
sequence objects that are still contracting.
In addition, measurements can be affected by extinction as well as excess emission from disk and/or accretion. 
The effective temperature is problematic for other reasons; it depends on atmosphere models and
can be altered by magnetic activity \citep{2012ApJ...756...47S}. This can lead us to underestimate
object masses by up to a factor of two.

For our chosen samples, accurate multi-band photometry is available, while
the spectroscopic follow-up is not complete. Therefore, we will rely
on photometry in the optical and near-infrared to estimate masses. We complement the 2MASS 
photometry provided by \citet{2009ApJS..184...18G} with optical photometry from 
\citet{1999ApJ...525..466L} and \citet{2003ApJ...593.1093L} for IC348 (Landolt R- and I-band) and from SONYC-I 
for NGC1333 (Sloan i- and z-band)\footnote{available from {\tt http://browndwarfs.org/sonyc}}. For some objects 
without 2MASS near-infrared magnitudes, we were able to complement the dataset using the photometry from 
\citet{2003AJ....125.2029M} for IC348\footnote{downloaded from\\ {\tt http://flamingos.astro.ufl.edu/sfsurvey/datarelease.html}} 
and SONYC-I for NGC1333. In total, the samples contain 142 (for IC348) and 95 (for NGC1333) objects with
photometry in JHK, with smaller subsets of 86 (IC348) and 23 (NGC1333) with additional
optical magnitudes available.

For the objects with 2MASS photometry, we also obtain the error as listed in the database (mostly 
between 0.03 and 0.05\,mag). Similarly, the photometry from \citet{2003AJ....125.2029M} provides errors
for all measurements. For the remaining photometry, errors for individual objects are not reported
in the literature. For the optical magnitudes in IC348, we adopt a generic and conservative uncertainty
of 0.1\,mag. For the SONYC magnitudes in NGC1333, we adopt errors of 0.1\,mag for the optical 
bands and 0.05\,mag for the near-infrared bands. 

Based on the information given in the papers listed above, the error values adopted for the optical
photometry should be typical for the samples. However, some objects might be affected by
additional uncertainties introduced to calibration imperfections. Since the z- and I-bands are located 
at the long-wavelength edge of the sensitivity of the optical CCDs, they are highly susceptible
to colour terms in the calibration, which are difficult to measure with the usual photometric standard
stars. This can introduce errors larger than 0.1\,mag in individual sources, which cannot be quantified accurately. 
This issue is a particular problem for very red sources, since most of their optical flux is emitted in 
the part of the spectrum where the CCD sensitivity declines.

We derive masses using three different sets of isochrones from the Lyon group, BT-Settl
(with AGSS2009 opacities), BT-Dusty (with AGSS2009 opacities), and BT-Nextgen (with GNC93 
opacities).\footnote{downloaded from\\ \tt{http://phoenix.ens-lyon.fr/simulator/index.faces}} 
The latter two are updated versions from the standard AMES-Dusty and Nextgen models. The main 
difference between the three sets is the treatment of dust. In contrast to Nextgen, Dusty
includes dust opacities. Settl includes a full dust cloud model. For more information on the isochrones, 
see \citet{2011ASPC..448...91A,2001ApJ...556..357A}. Note that atmospheric dust 
becomes a major source of opacity for $T_{\mathrm{eff}} \lesssim 2500$\,K \citep{2008MNRAS.391.1854H}, 
corresponding to $M \lesssim 0.02\,M_{\odot}$ for young brown dwarfs, which is the low-mass 
limit in our analysis, thus, the treatment of dust should not have a major effect on our results.
The isochrones predict absolute magnitudes 
as a function of object mass in all photometric bands for which observations are available. They 
cover the range from 0.02$\,M_{\odot}$ or below to 1.4\,$M_{\odot}$ and are available for ages 
starting from 1\,Myr, which is adequate for our purposes. 

To compare the observed with the predicted magnitudes, we first shift the isochrones from absolute
magnitude to the distance of our target regions, which enters here as a free parameter
(see Sect. \ref{s23}). We also re-bin the isochrones to a uniform stepsize of 
0.01$\,M_{\odot}$, using a linear interpolation over a small portion of the isochrone. We
then calculate a series of reddened isochrones for $A_V = 1-20$\,mag in steps of 1\,mag. For this
step, the choice of the extinction law is important (see Sect. \ref{s23}). The upper limit is chosen
to be 20\,mag, for two reasons. First, independent studies indicate that only a small fraction
in the Perseus star forming complex exceeds this extinction value \citep{2010A&A...512A..67L}.
Second, beyond this value the samples are biased towards bright sources, thus, incompleteness
becomes an issue.

\begin{figure*}
\center
\includegraphics[width=5.5cm,angle=-90]{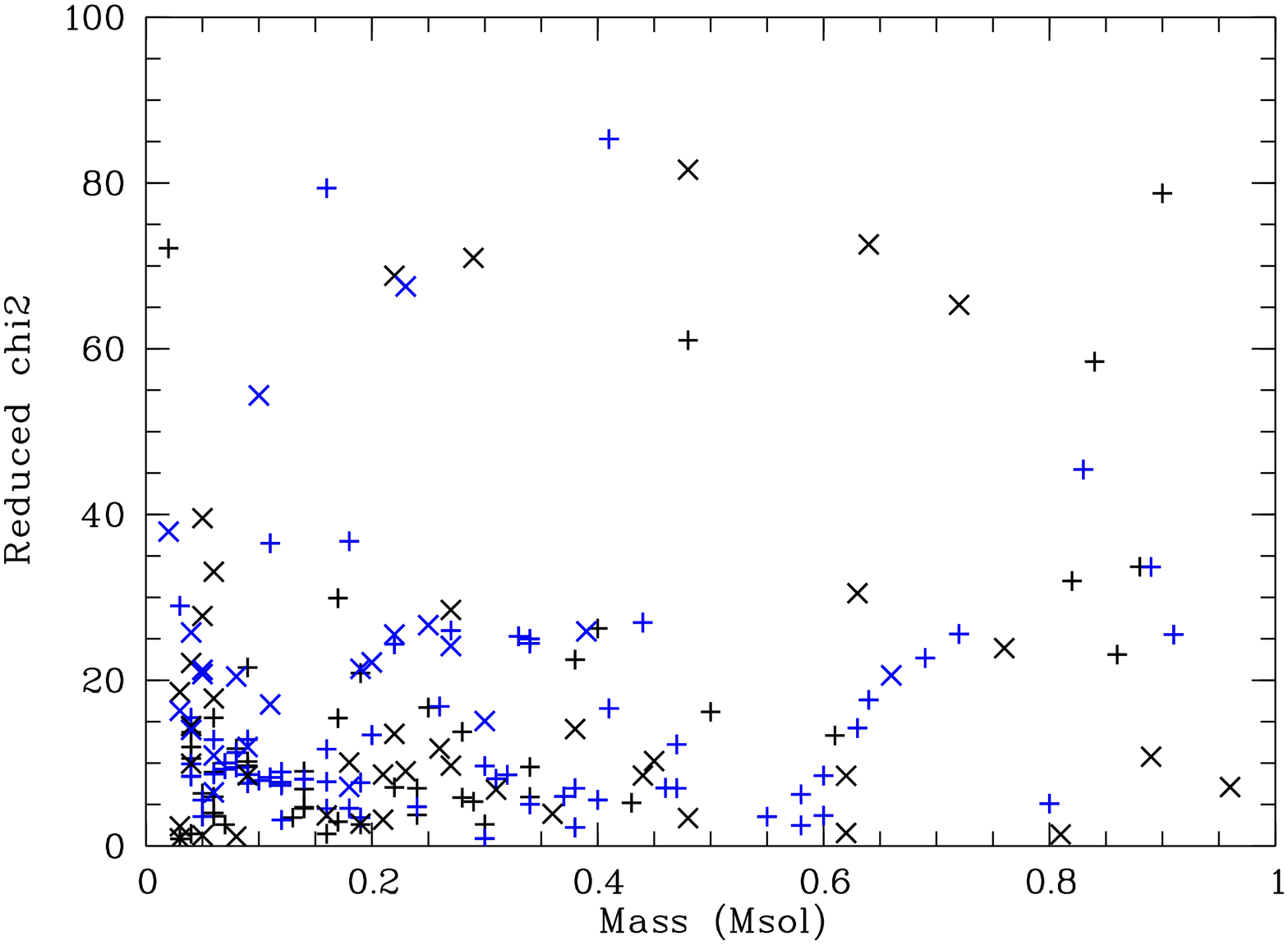} \hfill
\includegraphics[width=5.5cm,angle=-90]{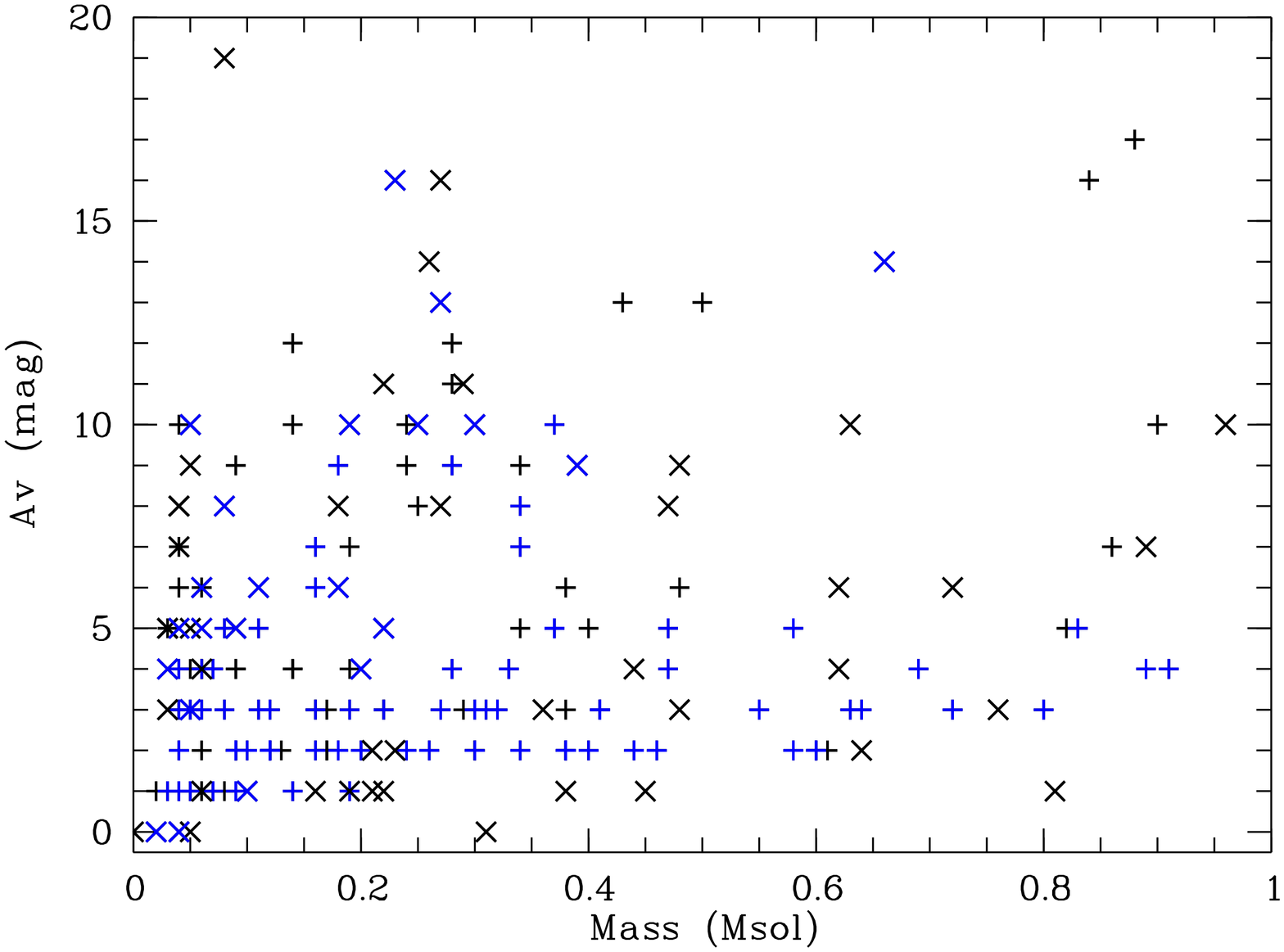} \hfill
\caption{Results from the fitting procedure for scenario \#1 (see Table \ref{t1}). Left panel: 
Reduced $\chi^2$ values vs. mass estimate, 6 additional datapoints with $\chi_r^2 >100$ are 
not plotted. Right panel: best fit $A_V$ vs. best fit mass estimate. Datapoints for IC348 are shown as 
plusses, those for NGC1333 as crosses. Objects with optical photometry are shown in blue. 
\label{f2}}
\end{figure*}

After these preparations, the best fit for mass and $A_V$ is determined with a $\chi^2$ minimization. 
The number of degrees of freedom in this process is the number of photometric bands for which data
is available ($N=3$ to $5$) minus the number of free parameters (mass and $A_V$, i.e. 2). 
For each object, we saved the combination of mass and $A_V$ that results in the minimum value for
$\chi^2$, the corresponding reduced $\chi^2$ ($\chi_r^2$, i.e. $\chi^2$ divided by the number of degrees
of freedom) and the number of available bands $N$. Objects with best fit value of $A_V =20$ are 
discarded from the analysis -- since this is the upper limit in our grid of isochrones, their 
mass estimate is not reliable. A typical example of the resulting $\chi_r^2$ plotted vs. the best mass
estimate is shown in Fig. \ref{f2}, left panel. This figure reveals that the procedure produces similar
fitting results for high- and low-mass objects and for objects with and without optical photometry.
We also show a typical $A_V$ vs. mass plot from this procedure (Fig. \ref{f2}, right panel). The
upper limit in $A_V$ is not changing significantly with mass, i.e. there is no evidence for an
extinction bias in these samples (apart from the $A_V<20$ cutoff).

The distribution of reduced $\chi^2$ can in principle be used to assess the goodness-of fit for our mass estimates.
For a good fit, we expect $\chi_r^2$ to have an average of 1.0 and a standard deviation of $\sim \sqrt{2/N}$.
However, as shown in Fig. \ref{f2}, our procedure yields significantly higher values in $\chi_r^2$. This
indicates that either the model does not reflect the data well or that the errors are underestimated.
In our case, the high values for $\chi_r^2$ are mostly explained by the fact that our model is discretely sampled 
in mass-$A_V$ space. The stepsize in mass and $A_V$ results in magnitude steps that are often larger than the typical 
photometric error. In addition, in some cases the errors of the optical photometry may be underestimated, see above. 
Therefore, we only use the procedure to select the best fit solution.

\subsection{The scenarios}
\label{s23}

In our estimation of masses from photometry, the distance, age, as well as the extinction
law, expressed in the quantity $R_V =  A_V / E_{(B-V)}$, are considered free parameters. In addition,
we have to choose the theoretical isochrone. What we call 'scenarios' in the following are combinations
of isochrone, distance, age, and extinction law for which we estimate masses. These scenarios have
been chosen to cover the plausible range of these parameters and to give insight into the impact of 
the specific choice of a parameter or an isochrone on the mass estimates. In the following, we 
justify the choice of the range of the parameters.

For the extinction, we use the parameterised law by \citet{1989ApJ...345..245C}, with $R_V = 3.1$, the 
canonical value used for the ISM. This law yields extinction offsets that are consistent with the often-used
extinction values published by \citet{1998ApJ...500..525S}. In reality, $R_V$ depends on the grain properties 
and is not the same for every line of sight; \citet{1989ApJ...345..245C} report values ranging from 2.6 to 5.6, 
with the overwhelming majority (22 out of 27 cases) below 4.5. We therefore use $R_V = 4.5$ as an alternative 
value to be able to assess the impact of the choice of $R_V$ on the mass estimates. 

The distances to the two clusters are not well constrained. The entire Perseus cloud is usually assumed 
to have an average distance of $\sim 300$\,pc, which we use as a default value. Based on the Hipparcos 
parallaxes for the early-type stars \citet{1999AJ....117..354D} estimate $318 \pm 27$\,pc for the cloud. 
Based on a kinematical analysis of a much larger sample of A stars, \citet{2002A&A...387..117B} infer 300\,pc 
(270-330\,pc). However, there are indications that NGC1333 is located at a shorter distance. 
\citet{2011PASJ...63....1H} report a distance of 235\,pc for NGC1333 based on interferometry of the 
maser emission from a source that may be associated with the cluster. This is also consistent with
an earlier photometric estimate of the distance of NGC1333 (220\,pc) by \citet{1990Ap&SS.166..315C}. In addition, 
\citet{2010A&A...512A..67L} suggest, based on extinction map analysis, that the northern part of the 
Perseus region (where IC348 is located) is slightly more distant than the southern part (where
NGC1333 is located). To take this into account, we use a distance of 230\,pc as an alternative value,
noting that this is only a viable option for NGC1333.

The ages that are typically quoted are 2-4\,Myr for IC348 and 1-3\,Myr for NGC1333 
\citep[see][and references therein]{2008hsf1.book..308B}. Judging from model-independent indicators of 
evolutionary state (fraction of objects with disks, fraction of objects in Class I stage, luminosity function), 
NGC1333 is definitely younger than IC348, and both are clearly younger than star forming regions with established 
ages of 5-10\,Myr like Upper Scorpius and the TW Hydrae Association 
\citep[e.g.][]{1996AJ....111.1964L,2001ApJ...553L.153H,2008ApJ...674..336G}.
In the context of our study, the relevant quantity is not the age of the cluster, but the average age
of the objects contained in our samples, which may not include the youngest, embedded population because
we require a near-infrared detection. In fact,
we showed in SONYC-IV that most of the very low mass objects in NGC1333 are consistent with an age
of 1-5\,Myr, based on their position in the Hertzsprung-Russell diagram. Therefore, we use a default value 
of 3\,Myr, which is plausible for both clusters. To assess the influence of the age on the mass estimates, 
we additionally estimate masses for an age of 1\,Myr.

To evaluate the impact of the choice of the parameters, we define 6 scenarios for which we estimate
object masses. These scenarios are listed in Table \ref{t1}. The default scenario \#1 uses a distance
of 300\,pc, an age of 3\,Myr, $R_V$ of 3.1, and the BT-Settl model, which has the most recent opacities
and the most advanced treatment of dust. In scenarios \#2 to \#4 we vary the cluster parameters. In 
scenario \#2 we use the younger age of 1\,Myr, in \#3 the alternative distance of 230\,pc, and in \#4 the
alternative value for $R_V$ of 4.5. In scenarios \#5 and \#6 we switch to the Nextgen and DUSTY
isochrones.

\begin{deluxetable}{lccclclcccc}
\tablecaption{Results from mass estimates for IC348 and NGC1333 \label{t1}}
\tablewidth{0pt}
\tablehead{
\colhead{cluster} & \colhead{no} & \colhead{D (pc)} & \colhead{t (Myr)} & \colhead{model} & \colhead{$R_V$} & \colhead{$R$} 
& \colhead{Min-Max\tablenotemark{a}} & Min-Max\tablenotemark{b}&\colhead{$\alpha$} & \colhead{$\overline{M}$}}
\tablecolumns{12}
\startdata
IC348 & 1 & 300 & 3 & BT-Settl   & 3.1 & 3.6 (96/27) & 2.8-4.4 & 2.6-4.1 & 0.72 & 0.27\\
IC348 & 2 & 300 & 1 & BT-Settl   & 3.1 & 2.1 (88/42) & 1.7-2.5 & 1.5-2.9 & 0.95 & 0.13\\
IC348 & 3 & 230 & 3 & BT-Settl   & 3.1 & 1.9 (84/45) & 1.5-2.2 & 1.3-2.1 & 1.00 & 0.16\\
IC348 & 4 & 300 & 3 & BT-Settl   & 4.5 & 3.2 (94/29) & 2.6-4.0 & 2.7-4.3 & 0.76 & 0.26\\
IC348 & 5 & 300 & 3 & BT-Nextgen & 3.1 & 4.0 (99/25) & 3.1-4.9 & 3.4-5.5 & 0.67 & 0.19\\
IC348 & 6 & 300 & 3 & BT-Dusty   & 3.1 & 2.9 (92/32) & 2.3-3.5 & 2.4-3.4 & 0.81 & 0.22\\
\noalign{\smallskip}
\hline
\noalign{\smallskip}
N1333 & 1 & 300 & 3 & BT-Settl   & 3.1 & 2.2 (43/20) & 1.6-2.8 & 1.9-2.5 & 0.94 & 0.27\\
N1333 & 2 & 300 & 1 & BT-Settl   & 3.1 & 2.1 (47/22) & 1.6-2.7 & 2.0-2.5 & 0.94 & 0.13\\
N1333 & 3 & 230 & 3 & BT-Settl   & 3.1 & 2.4 (47/20) & 1.8-3.0 & 2.2-2.4 & 0.90 & 0.18\\
N1333 & 4 & 300 & 3 & BT-Settl   & 4.5 & 2.0 (43/21) & 1.6-2.6 & 1.9-2.5 & 0.96 & 0.30\\
N1333 & 5 & 300 & 3 & BT-Nextgen & 3.1 & 2.2 (44/20) & 1.7-2.8 & 1.9-2.6 & 0.93 & 0.18\\
N1333 & 6 & 300 & 3 & BT-Dusty   & 3.1 & 1.9 (42/22) & 1.5-2.5 & 1.8-2.0 & 0.99 & 0.22\\
\enddata
\tablenotetext{a}{Statistical uncertainties for $R$, only dependent on sample size}
\tablenotetext{b}{Min-max range in $R$ based on the number of objects close to the substellar limit, see Sect. \ref{s33}}
\end{deluxetable}

\begin{figure*}
\center
\includegraphics[width=5.5cm,angle=-90]{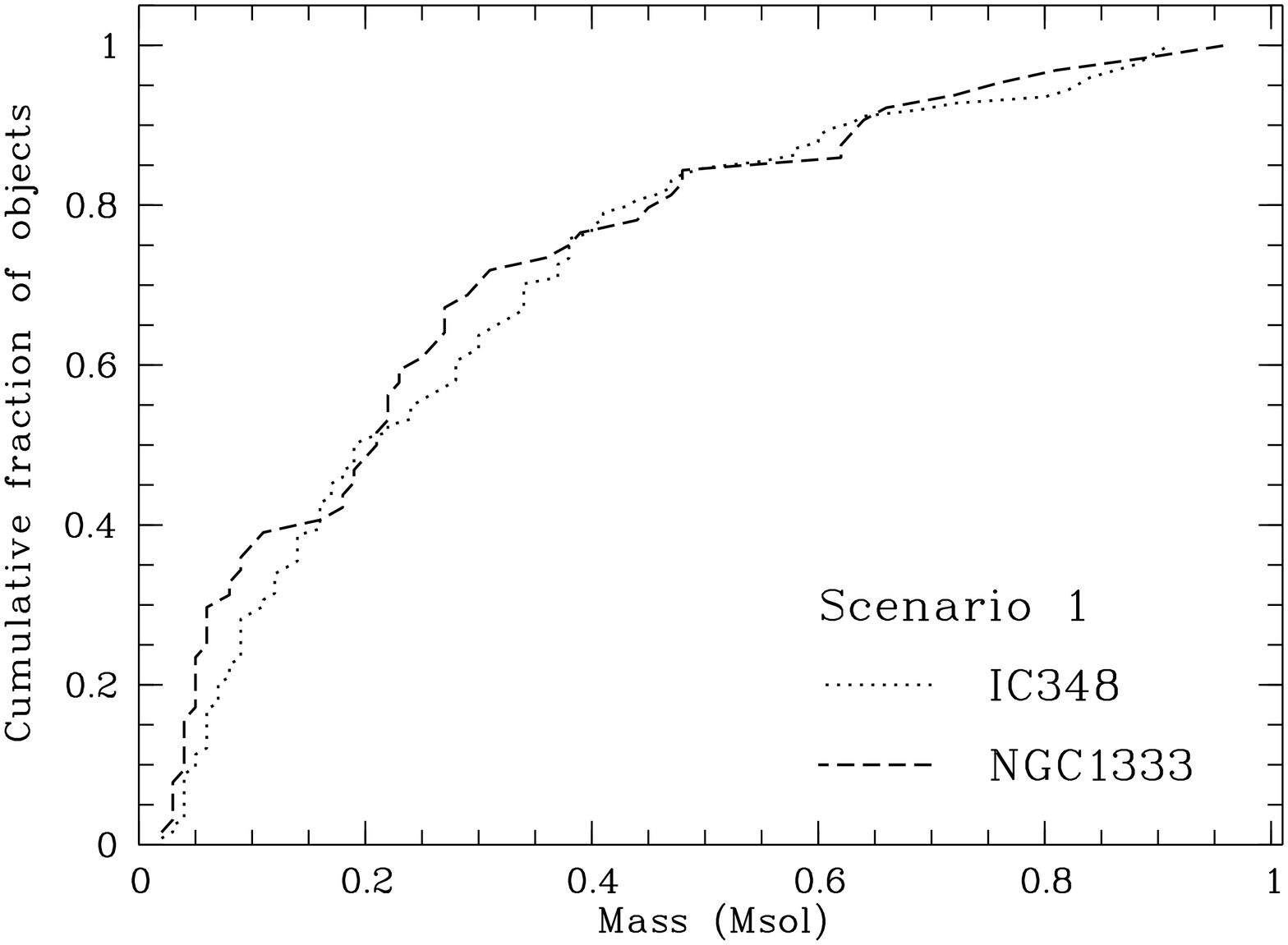} \hfill
\includegraphics[width=5.5cm,angle=-90]{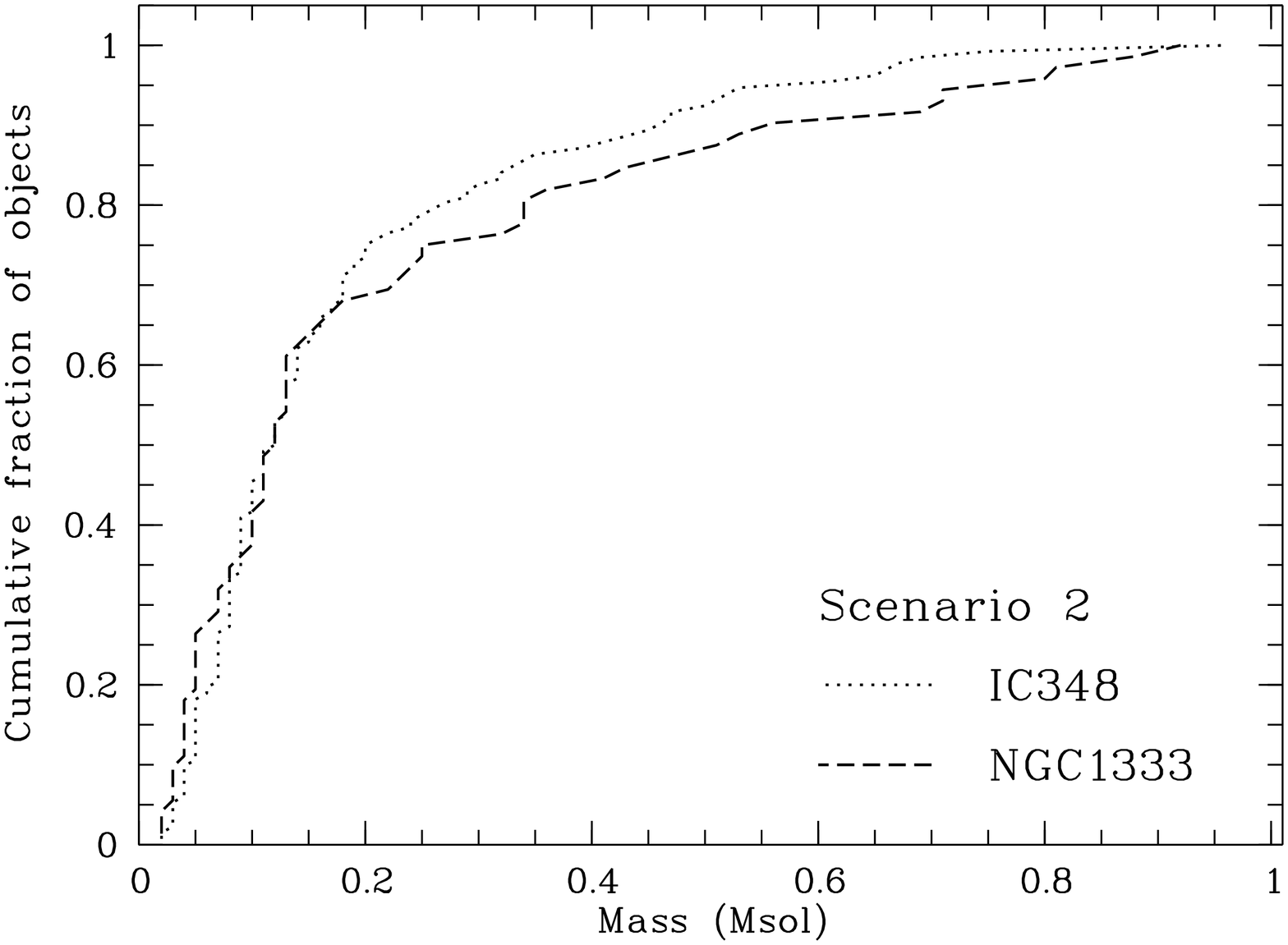} \\
\includegraphics[width=5.5cm,angle=-90]{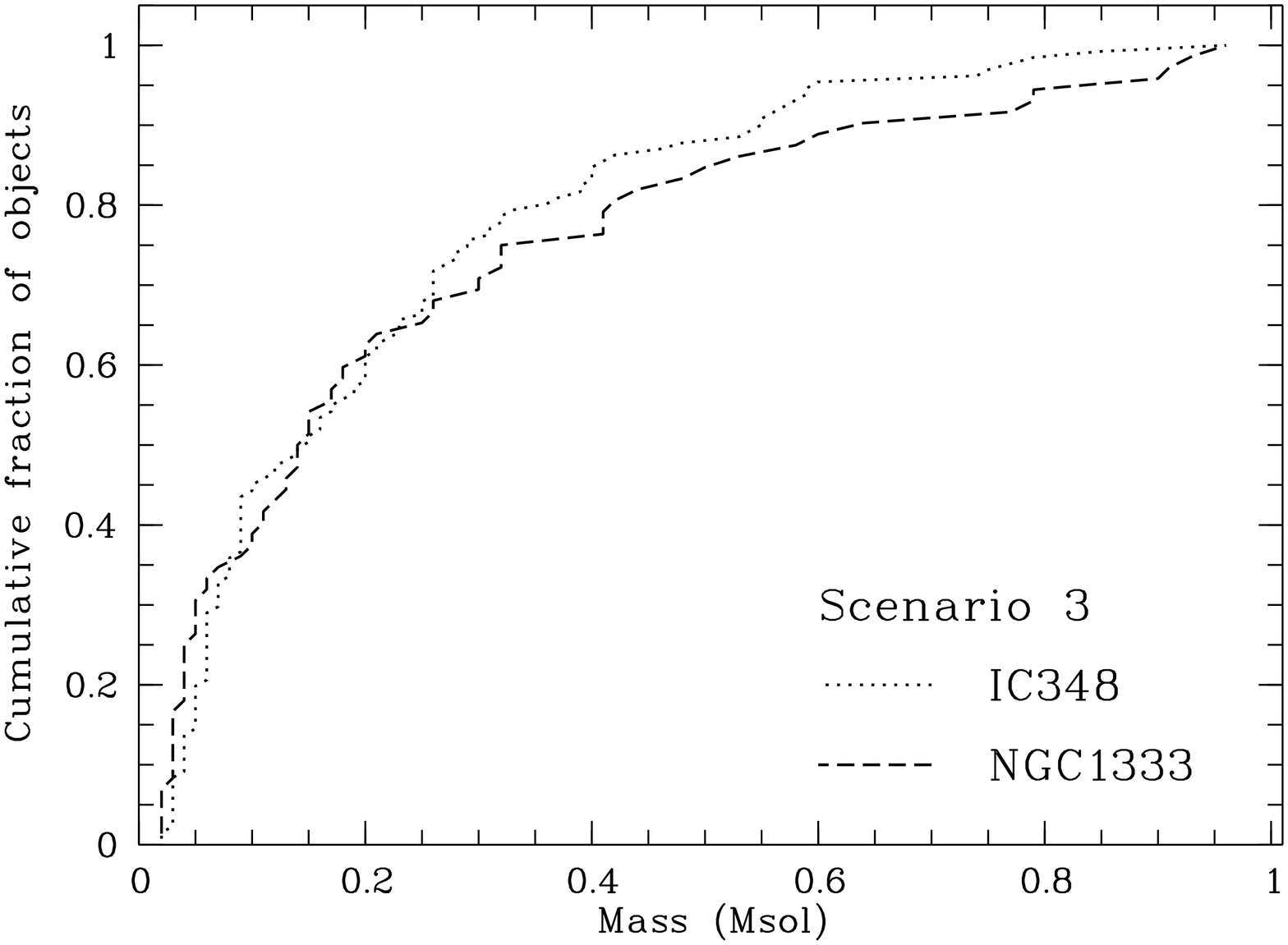} \hfill
\includegraphics[width=5.5cm,angle=-90]{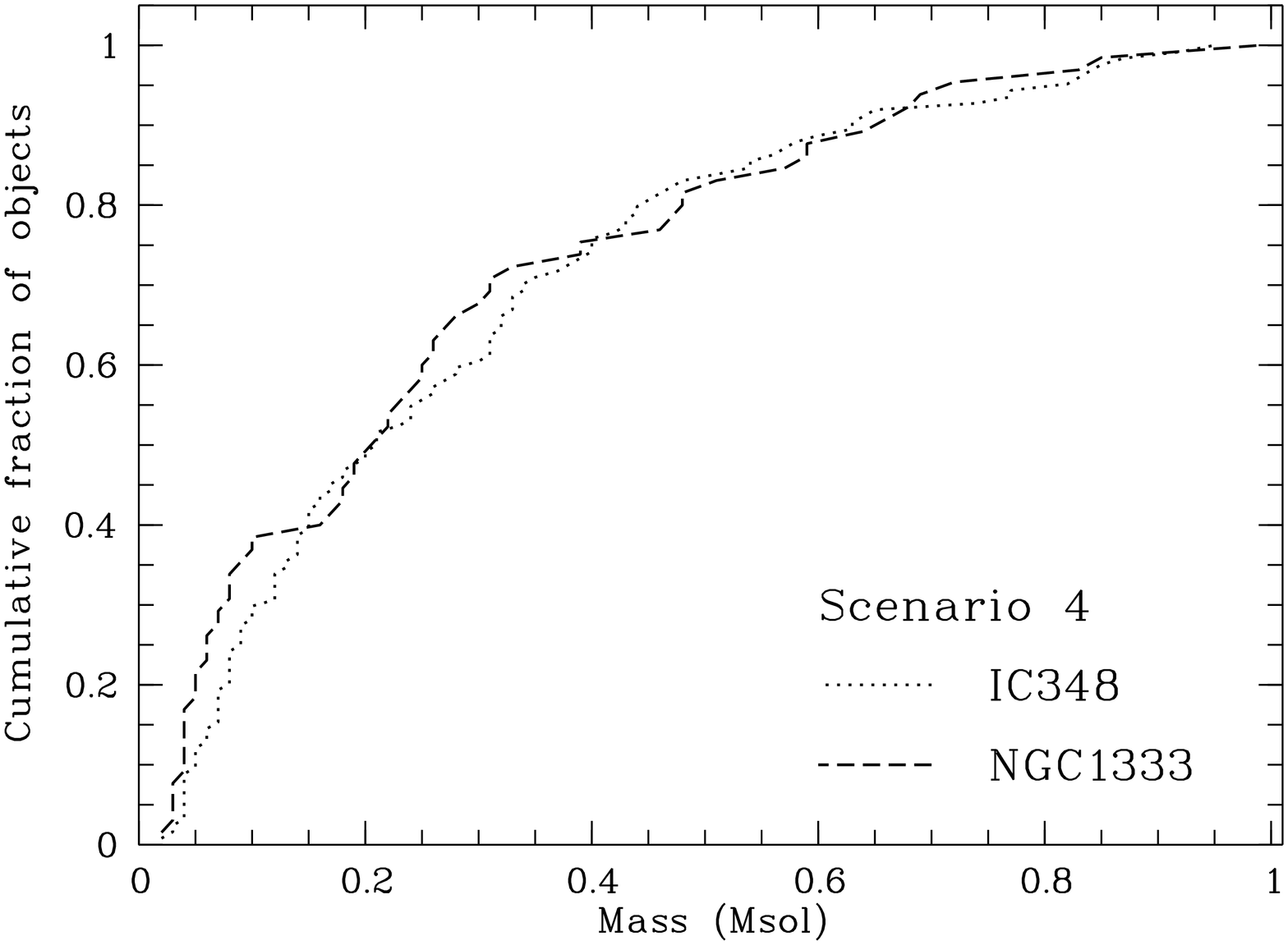} \\
\includegraphics[width=5.5cm,angle=-90]{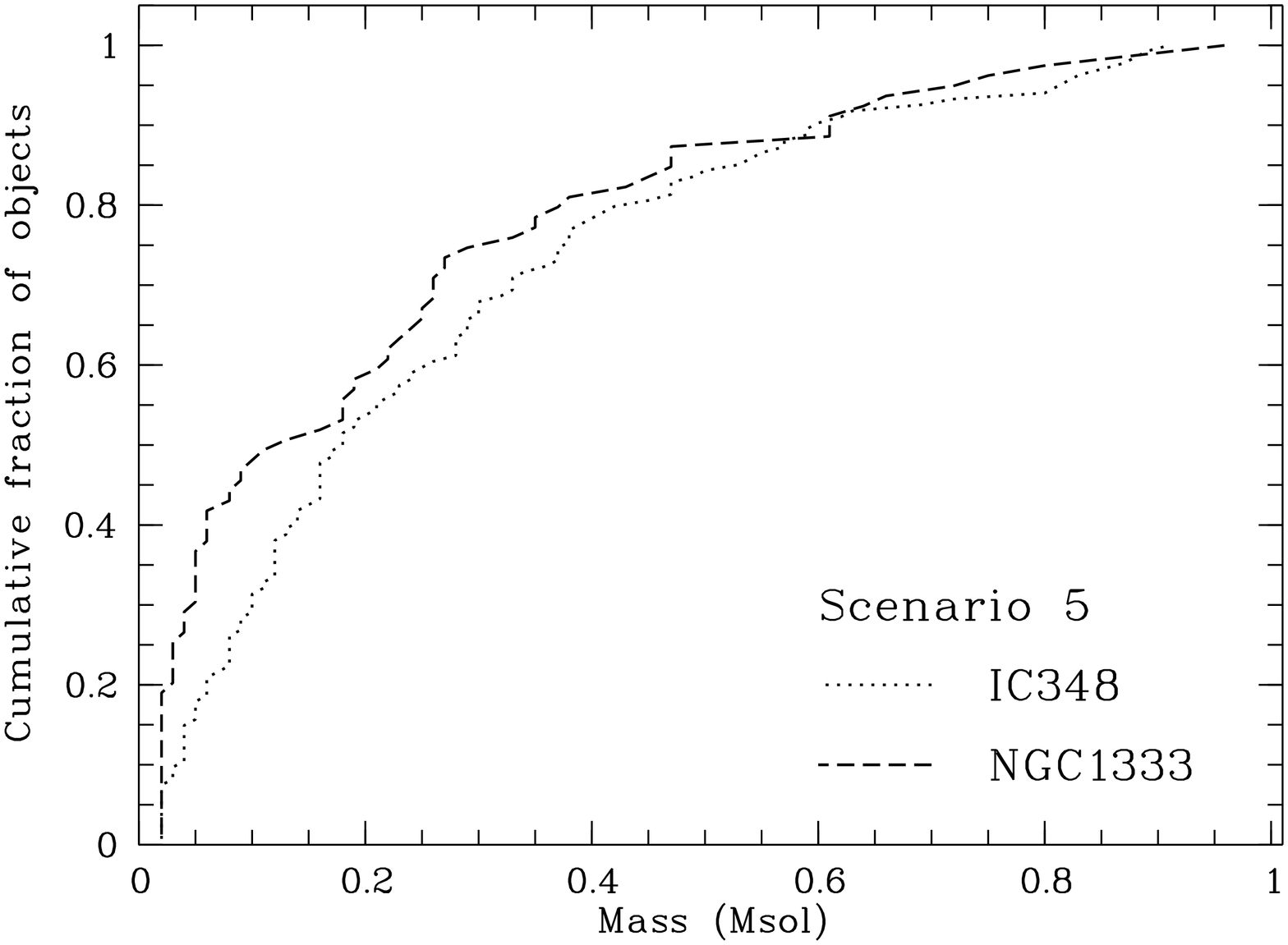} \hfill
\includegraphics[width=5.5cm,angle=-90]{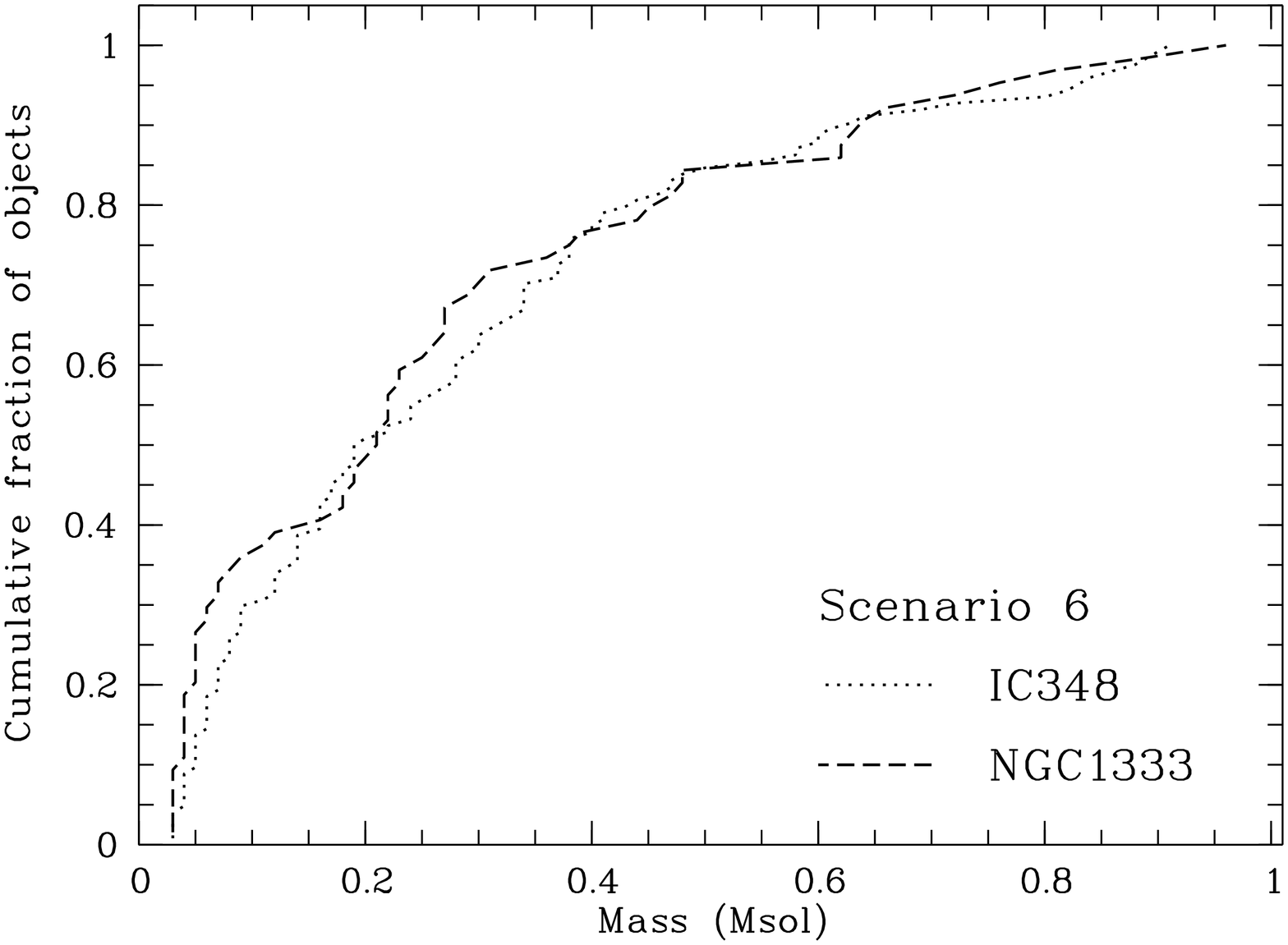} \\
\caption{Cumulative distribution of objects as a function of mass (i.e. the fraction of objects below
a given mass). Shown are the results for all 6 scenarios from Table \ref{t1}. The parent sample for these
plots has $A_V<20$ and $M\le 1.0\,M_{\odot}$.
\label{f1}}
\end{figure*}

From the resulting mass distribution in a given scenario, we derive the cumulative distribution of object 
masses, i.e. the fraction of objects below a given mass, as a function of mass. These plots are shown in Fig.
\ref{f1}. For each scenario, we determine the number of objects with 
$0.03\le M \le 0.08\,M_{\odot}$ (BDs) and with $0.08< M <1.0\,M_{\odot}$ (stars) and calculate the 
star/BD ratio $R$. To be able to compare with the literature, we also calculate the slope of the 
mass function $\alpha$ for the power-law parameterisation $dN/dM \propto M^{-\alpha}$, directly from 
the star/BD ratios, i.e. using two bins in mass, one for BDs and one for stars. In addition, we 
derive the median mass $\overline{M}$ for each scenario. The resulting parameters for the 6 
scenarios are given in Table \ref{t1}.

\subsection{Error budget}
\label{s33}

An important part of this work is to evaluate the errors in the derived parameters. In our chosen experiment,
five factors contribute to the uncertainties:

{\bf 1) Sample size:} The part that is easiest to quantify is the statistical uncertainty, which is purely determined 
by the sample size. For this paper, we use the same approach as in SONYC-IV, which is based on the IDL scripts presented
in \citet{2011PASA...28..128C}. In short, we calculate Bayesian confidence intervals from the beta distribution. We note 
that for large samples this procedure gives results that are very similar to binomial confidence intervals. The
resulting values are listed in Table \ref{t1}, column 8. Typically, the sample size introduces an error in $R$ of about
$\pm$0.3-0.9 for IC348 and $\pm$0.5-0.6 for NGC1333.

{\bf 2) Models:} Masses are only defined in relation to evolutionary models, and at young ages the deficiencies of the 
available tracks are well documented \citep{2002A&A...382..563B,2003A&A...398.1081W}. However, at the moment no tracks 
with self-consistent treatment of the collapse and infall are available to the community.\footnote{If, in a hypothetical 
future, evolutionary tracks are linked with the initial conditions via realistic models 
for collapse, infall, and dynamical evolution in clusters, one would directly compare the predicted with the observed 
luminosity functions or HRDs. The semi-empirical estimate of IMFs for star forming regions, as it is done in this paper 
and in many others in the literature, would become obsolete.} Based on a dynamical mass
estimate for the very low mass pre-main sequence object AB Dor C, which is older than our target regions, 
\citet{2005Natur.433..286C} claim that the existing mass-luminosity relations underestimate masses by a factor of 
about two for young objects, but this claim has been questioned \citep{2006ApJ...638..887L}. At young ages and very 
low masses, the only direct benchmark test for these tracks is the eclipsing brown dwarf binary 2M J05352184-0546085 
\citep{2007ApJ...664.1154S}. The Baraffe et al. isochrones fail to reproduce the surprising temperature reversal in 
this object (i.e. the more massive object is cooler than the secondary), an effect that is likely related to the 
presence of strong magnetic fields on the primary \citep{2012ApJ...756...47S}. The luminosities of the two components, 
however, are consistent with the isochrones. \citet{2007MNRAS.380..541I} have discovered and analysed a very young 
eclipsing binary with component masses around $\sim 0.2\,M_{\odot}$; that system confirms the isochrones within the 
errorbars as well. Thus, while more work is required to calibrate the isochrones, some preliminary 
trust in their validity seems warranted.

{\bf 3) Cluster parameters:} As discussed in Sect. \ref{s22}, several properties of the clusters affect the mass 
estimates, in particular the distance, the age, and the extinction law. From our set of scenarios documented in Table 
\ref{t1} (scenarios 1-4) we can assess how the uncertainties in these parameters propagate through the procedure. 
In general, changes in age and distance cause significant changes in the estimated mass distribution, while a change 
in the extinction does not. For NGC1333, the induced variations in the star/BD ratios are small; $R$ varies from 2.0 
to 2.4, smaller than
the statistical uncertainties. For IC348, the scatter is larger, from 1.9 to 3.6, but excluding the implausible
scenarios with age of 1\,Myr and distance of 230\,pc this range shrinks to 3.2-3.6. We note that the star/BD ratio 
increases somewhat with assumed age. For example, for IC348 and an (unlikely) age of 5\,Myr we obtain $R = 6.7$. 
This is easy to understand -- as the objects evolve, they become fainter i.e. the same magnitude corresponds to a 
larger mass. As a result, objects move from the substellar to the stellar domain, and the star/BD ratio increases.

{\bf 4) Completeness:} The samples we are using have the same depth and completeness, in terms of magnitudes, which
eliminates one major source of uncertainty. However, in terms of object masses the depth of the survey depends on
the assumed distance and age. Assuming a shorter distance, as well as a younger age, will produce lower masses for the
same magnitudes, i.e. the entire mass distribution would be shifted to lower masses, including the limits for
completeness. For our default scenario with age of 3\,Myr and distance of 300\,pc, the magnitude limit of the samples 
correponds to $\sim 0.02\,M_{\odot}$. The alternative distance of 230\,pc 
(scenario \#3) implies a magnitude shift by 0.6\,mag. According to BT-Settl isochrones, this translates to a mass limit of
0.015$\,M_{\odot}$. The alternative age of 1\,Myr (scenario \#2) yields a new mass limit of 0.012-0.015$\,M_{\odot}$.
Thus, in these scenarios we would be sensitive to slightly lower mass objects, but since the object density is
very low in this mass domain, this has only a minuscule effect on the mass distibution. Furthermore, it is not going 
to affect the star/BD ratios.

{\bf 5) Degeneracy:} In many cases there are multiple mass/$A_V$ combinations that fit the data with
a similar $\chi^2$. In particular, for an object with a best mass estimate around the Hydrogen burning limit 
(0.07-0.09$\,M_{\odot}$) this method is unable to distinguish between a star and a BD. As an estimate 
of the introduced uncertainty in $R$ we selected all objects in this mass regime and included them first in the BD count 
(for the lower limit of the star/BD ratio) and then in the star count (for the upper limit). The resulting ranges 
in $R$ are listed in Table \ref{t1}, column 9. These intervals are $\pm$0.1-0.3 for NGC1333 and $\pm$0.5-1.0 for IC348.
We note that the degeneracy is less of a problem for the two other mass thresholds involved in the calculation of 
$R$ (0.03 and 1.0$\,M_{\odot}$), simply because the number of objects around these limits is low.

Combining these error sources, the values of $R$ are affected by an uncertainty of approximately $\pm 1$ for the two 
samples studied here. Currently this is about the best that can be done in terms of estimating
this indicator for young star forming regions. This translates to an uncertainty of about $\pm 0.1$ in the power-law
slope $\alpha$. The median of the mass function can be estimated with an accuracy of $\pm 0.1\,M_{\odot}$ for nearby
star forming regions.

Looking ahead, there are obvious ways to lower these 
uncertainties. First, future evolutionary tracks need to include realistic initial conditions and require more 
detailed calibration (e.g., with eclipsing binaries). Second, the extinction parameters 
need to be studied in more detail for individual regions. Third, independent estimates for the (relative or absolute) 
ages of young clusters are needed. Fourth, the accuracy in the distances of these clusters (and many other well-studied 
star forming regions) should be improved. In this respect, the Gaia satellite will be a major opportunity, as it is
anticipated to be able to measure distances with 1\% accuracy for open clusters within 1\,kpc \citep{2011sca..conf...11P}, 
a huge improvement over the current estimates. Fifth, it is a worthwhile goal to obtain comprehensive sets of 
multi-filter photometry for star forming regions. And sixth, additional survey work in rich star forming regions with 
significantly more members (factor 10 or more) than the well-studied nearby regions can be used to minimize the 
statistical uncertainties. However, since these regions are only found at large distances of $>2$\,kpc, such studies
have to be postponed until larger facilities, namely JWST or ELTs, are available.

\section{Results}
\label{s3}

\subsection{The cumulative distribution}
\label{s35}

Our procedure yields for each scenario and for each cluster a distribution of masses. Prior to calculating parameters for 
the IMF, we examine these distributions directly.
In Fig. \ref{f1} we show for all 6 scenarios the cumulative distribution of object masses, i.e. the fraction of
objects below a given mass, as a function of mass. We compare the 36 combinations of the functions shown in Fig. 
\ref{f1} (6 for IC348 and 6 for NGC1333) with a Kolmogorov-Smirnov (KS) test to search for differences in the mass 
distribution. By doing that, we test the null hypothesis that two given distributions are drawn from the same 
parent distribution. In Table \ref{t2} we list the probabilities that this hypothesis is valid.

\begin{deluxetable}{lccccccc}
\tablecaption{Probability that two cumulative distributions of object masses are
drawn from the same parent distribution. Combinations with significant differences are marked. \label{t2}}
\tablewidth{0pt}
\tablehead{
\colhead{Scenario} & \colhead{N1333-1} & \colhead{N1333-2} & \colhead{N1333-3} & \colhead{N1333-4} & 
\colhead{N1333-5} & \colhead{N1333-6} \\
\colhead{} & \colhead{default} & \colhead{$t=1$\,Myr} & \colhead{$D=230$\,pc} & \colhead{$R_V = 4.5$} & 
\colhead{Nextgen} & \colhead{Dusty}}
\tablecolumns{8}
\startdata
IC348-1: default     & 0.26        & {\bf 0.02}  & {\bf 0.04}  & 0.50        & {\bf 0.001} & 0.17        \\
IC348-2: $t=1$\,Myr  & {\bf 0.002} & 0.49        & 0.10        & {\bf 0.003} & {\bf 0.01}  & {\bf 0.002} \\
IC348-3: $D=230$\,pc & 0.32        & 0.33        & 0.32        & 0.39	& {\bf 0.03}  & 0.32	    \\
IC348-4: $R_V = 4.5$ & 0.26        & {\bf 0.003} & 0.05        & 0.13	& {\bf 0.001} & 0.17	    \\
IC348-5: Nextgen     & 0.80        & {\bf 0.016} & 0.30        & 0.76	& {\bf 0.02}  & 0.54        \\
IC348-6: Dusty       & 0.47        & {\bf 0.02}  & 0.08        & 0.74	& {\bf 0.004} & 0.32        \\
\enddata
\end{deluxetable}

For 14 out of 36 combinations, this probability is $<5$\%, i.e. the null hypothesis should be rejected. 6 of 
them are combinations that include scenario \#5 for NGC1333, i.e. the one that uses the Nextgen isochrone. 
This scenario produces an unusual large number of low-mass BDs (0.02-0.05$\,M_{\odot}$) for NGC1333, which causes
the discrepancy with other distributions. This shows that apparent differences in the mass distribution of 
young clusters can be introduced simply by the choice of the isochrone. We caution against comparing mass
distributions derived with inconsistent isochrones.

\begin{figure*}
\center
\includegraphics[width=5.5cm,angle=-90]{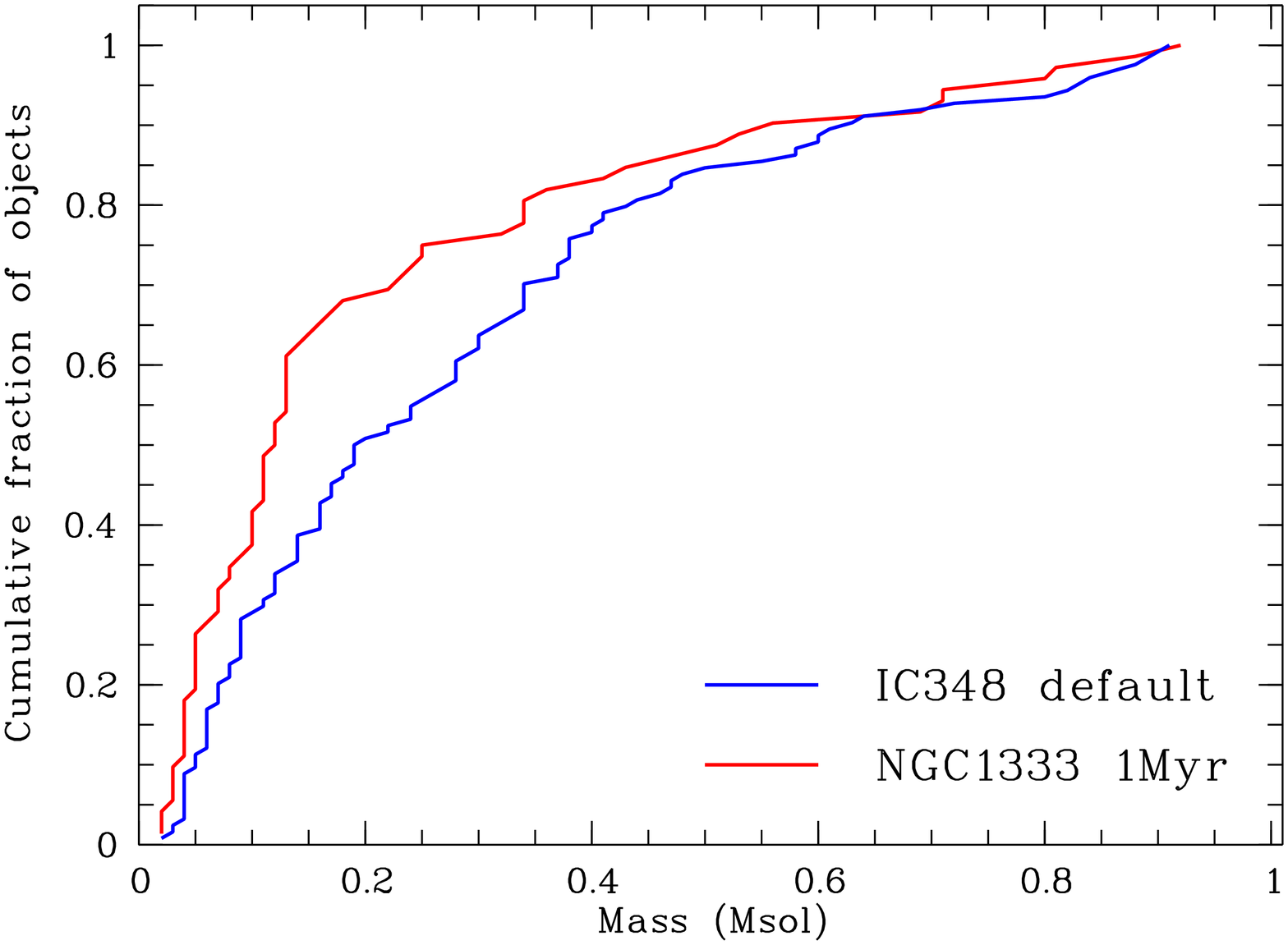} \hfill
\includegraphics[width=5.5cm,angle=-90]{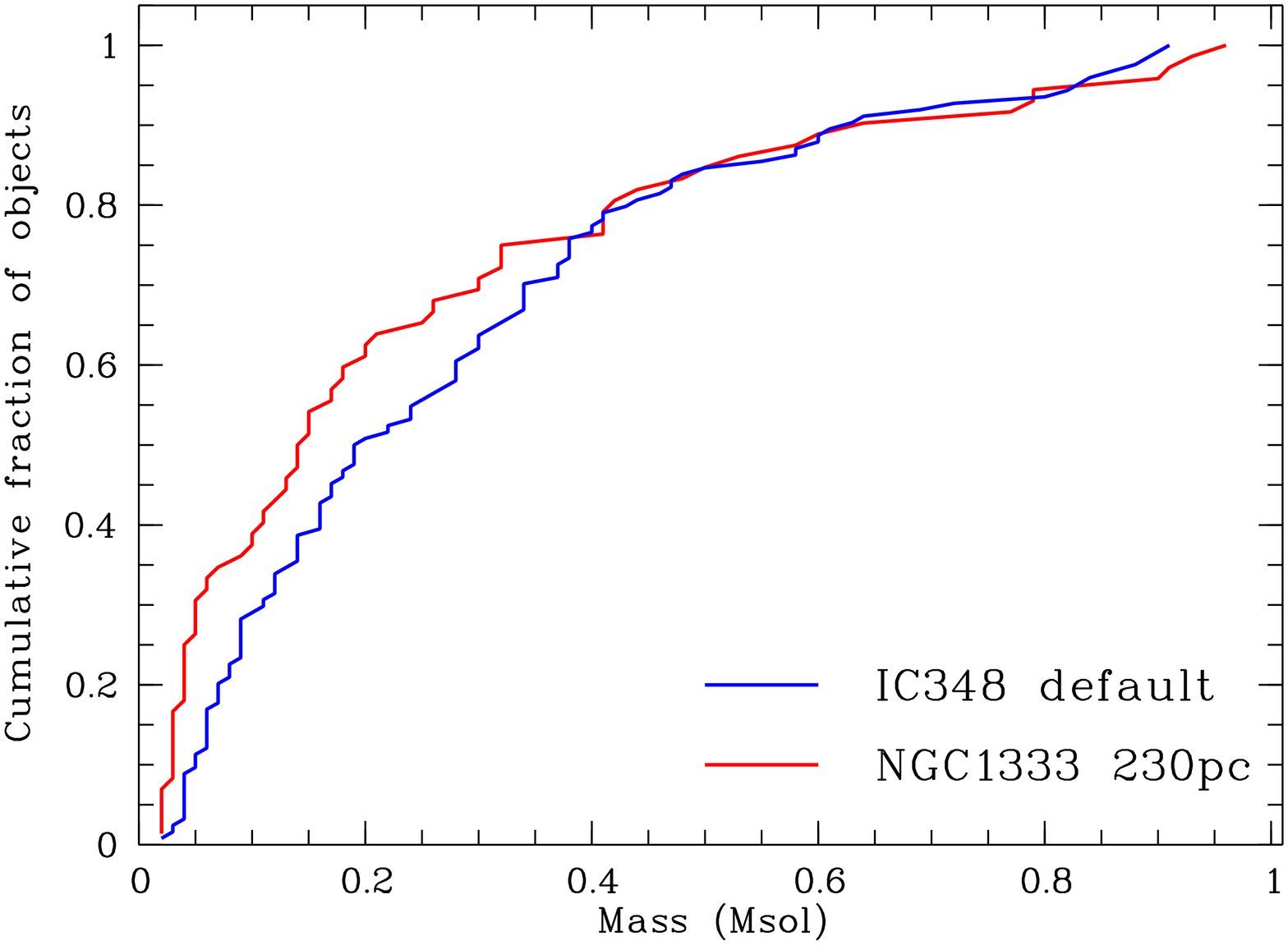} \\
\caption{Cumulative distribution of objects as a function of mass (i.e. the fraction of objects below
a given mass). Shown are two cases for which the distribution in IC348 is significantly different from the 
one in NGC1333. The parent sample for these plots has $A_V<20$ and $M\le 1.0\,M_{\odot}$. In both cases
the default scenario \#1 is used for IC348. In the left panel, we use scenario \#2 for NGC1333,
which assumes an age of 1\,Myr. In the right panel, we use scenario \#3 for NGC1333, which assumes a 
distance of 230\,pc. 
\label{f3}}
\end{figure*}

7 further combinations of scenarios with significant differences in the mass distribution include scenario
\#2 for one of the clusters, i.e. the scenario with an age of 1\,Myr. As an example, we show in Fig. \ref{f3} (left
panel) a comparison between the default scenario for IC348 and scenario \#2 for NGC1333. From this figure the
origin of the difference is clear -- the mass distribution for NGC1333 shows a pronounced 'knee' at 0.15$\,M_{\odot}$. 
However, the mass distribution in IC348 gives almost exactly the same 'knee' when assuming an age of 1\,Myr
(upper left panel in Fig. \ref{f1}). This effect is best explained by the 'knee' in the mass-luminosity 
relation at this age, which is predicted to become weaker with age. The resulting differences in the mass 
distribution cannot be attributed to environmental differences. 

The combination of the default scenario for IC348 and scenario \#3 for NGC1333 produces a significant difference 
as well (see Fig. \ref{f3}, right panel). This combination assumes the shorter distance of 230\,pc for NGC1333. We 
note that three further combinations with this assumption give marginally significant differences with probabilities 
between 5-10\%. As explained in Sect. \ref{s23}, multiple independent studies indicate that NGC1333 is located
at a shorter distance than IC348, thus, this scenario is plausible.

In summary, his comparison shows that the choice of the cluster parameters and the choice of the isochrone can
have noticable effects on the estimated distribution of object masses. We find that there could be significant 
differences in the mass distributions between the two clusters, if NGC1333 is at a shorter distance than IC348. 
In this case, our analysis indicates a larger proportion of very low mass objects with masses 
$<0.3\,M_{\odot}$ in NGC1333. 

\subsection{The star/BD ratio}
\label{s34}

From the mass distributions in the 6 scenarios we calculated the star/BD ratio, see Table \ref{t1} for
the results. For the cluster NGC1333, our 6 scenarios give star/BD ratios of 1.9-2.4, which is a range 
comparable to the statistical uncertainties (see Sect. \ref{s33}). This means that our previous estimate 
for this cluster from SONYC-IV ($R\sim 2.3$) is confirmed. 

For IC348, we find a wider range of values between 1.9 and 4.0. These values imply that 
star/BD ratios reported in the literature for IC348 of $\sim 8$ \citep{2003ApJ...593.1093L,2008ApJ...683L.183A} 
are overestimated. A possible reason for these large values is survey incompleteness or low number statistics.
As explained in Sect. \ref{s33}, the star/BD ratio depends on age, therefore these large star/BD ratios in the 
literature would become viable if IC348 is in fact significantly older than 3\,Myr. This is unlikely, as 
age-dependent observable quantities like the disk fraction are comparable to other 2-3\,Myr old clusters 
\citep{2013MNRAS.429..903D}. 

For IC348 an age of 1\,Myr and a distance of 230\,pc are not plausible (Sect. \ref{s23}). Excluding the scenarios
using these parameters gives a range for $R$ between 2.9 and 4.0, which is our best estimate for this cluster. 
These values are somewhat larger than in NGC1333, although still within the margin of error. For example, for the 
default case the star/BD ratio is 3.6 for IC348 with a lower limit of 2.6, and 2.2 for NGC1333 with an upper limit 
of 2.5. This is before taking into account the statistical uncertainties. Thus, based on the star/BD ratio the 
evidence for regional differences in the mass distribution of IC348 and NGC1333 is tentative.

\subsection{Other parameters}

In Table \ref{t1} we also report two other quantities that are used in the literature to describe the IMF. 
The power-law slope of the mass function $\alpha$ is directly determined from the star/BD ratio and thus reflects the same
trends reported in Sect. \ref{s34}. For NGC1333, $\alpha$ is 0.9-1.0; for IC349 0.7-1.0, or 0.7-0.8 after excluding
the implausible scenarios. Note that the slope is 
an average value for the mass range 0.03 to 1.0$\,M_{\odot}$; therefore it is not unexpected to find values 
somewhere between the tyipcal slope of 1.3 in the regime of low-mass stars \citep{2001MNRAS.322..231K} and the typical 
value of $\sim 0.6$ in the very low mass regime (see SONYC-IV).

Independent from the star/BD ratio, we determine the median mass for each scenario. These values vary between
0.13 and 0.30$\,M_{\odot}$, again indicating that the choice of the cluster parameters and the choice of the 
isochrone affect the results considerably. For a given scenario, the two clusters have very similar median masses 
(maximum difference is 0.04$\,M_{\odot}$ for scenario 4). Increasing age and distance will also increase 
the median mass. Note that \citet{2013A&A...549A.123A} have recently determined the mass function for IC348 from 
a different survey and find a characteristic mass (in the lognormal mass function) of 0.21-0.22$\,M_{\odot}$ 
consistent with the values derived here.

\section{Implications for brown dwarf formation}

In Sect. \ref{s3} we established that differences in the mass distributions of the two clusters are significant,
if NGC1333 is closer than IC348. In addition, there is tentative evidence that the star/BD ratio in
IC348 is slightly larger than in NGC1333. Given that our two target regions differ in object density by a factor
of 4-7 (Sect. \ref{s22}), this finding can in principle be used to put constraints on theories for BD formation in 
which the stellar density is a critical parameter for the yield.

One popular scenario to form BDs is as part of dynamical cluster formation. Here, very low mass objects are
removed from their accretion reservoir by dynamical ejections and thus stop their growth; the final mass is 
set by the competition between accretion and ejection \citep{2012MNRAS.419.3115B}. In this model the efficiency 
of BD formation is partly controlled by the likelihood for dynamical encounters which is related to the object 
density. The most recent radiation-hydrodynamical simulations by Bate (see their Table 1) are comparable to 
the clusters studied here in terms of initial cloud mass and number of objects produced. The simulations yield
star/BD ratios ($>2.6$, $>4.1$) and median masses (0.21, 0.24$\,M_{\odot}$) that are consistent with our 
empirical results. However, the impact of object density is difficult to judge, since the simulations have only 
been carried out for a very limited set of initial conditions.

Another way to form BDs that has been suggested in the literature is gravitational fragmentation of infalling 
gas into a stellar cluster \citep{2008MNRAS.389.1556B}. Here the potential well, and thus the object density 
in the cluster, is a critical parameter for the efficiency of BD formation. The BDs and very low mass objects
are expected to be formed preferentially in regions with high stellar density. Qualitatively the predictions 
from this scenario are confirmed by our analysis: Under plausible assumptions, the denser cluster NGC1333 has 
indeed a larger fraction of very low mass stars and brown dwarfs (see Sect. \ref{s34}). 

Their figure 7 shows the BD fraction as a function of object density from their simulations. The two regions
investigated in the current paper are both at the low end of the considered densities (1-100\,pc$^{-3}$). For these
densities, the predicted BD fractions are between 7 and 13\%. In their paper the BD fraction is calculated as
the number of BDs divided by the total number of objects. From our mass distributions, this quantity 
is $\sim 20$\% in IC348 and $\sim 30$\% in NGC1333, i.e. the predicted values are lower than the observed one.
If this formation mechanism plays a role and the predictions are realistic, it could only contribute about one third 
to half of the BDs in the clusters. Other mechanisms, for example disk fragmentation followed by the ejection
of embryonic or proto-brown dwarfs \citep{2009MNRAS.392..413S,2012ApJ...750...30B}, could contribute to the final 
tally of substellar objects in the young clusters.

Judged by their figure 7, an increase in the object density by one order of magnitudes would result in an increase 
in the BD fraction by a factor of about 2. This is consistent with the observed difference between IC348 and NGC1333,
although such a difference is, as explained in Sect. \ref{s34}, still within the uncertainties. Therefore, the 
scenario remains viable, but cannot be rigorously verified with the current surveys. An important test for the
theory would be the measurement of the BD fraction in a cluster that is significantly denser than NGC1333, such
as RCW38 \citep{2008hsf2.book..124W} or the Orion Nebula Cluster. So far, the survey results in the ONC give
inconsistent answers regarding the frequency of very low mass objects 
\citep{2011A&A...534A..10A,2012ApJ...748...14D}. 

The aforementioned scenario for brown dwarf formation via disk fragmentation could also result in a star/BD ratio
that depends on stellar density, if some of the fragmentation processes are driven by stellar encounters 
\citep{2010ApJ...717..577T} or disk-disk collisions \citep{2010MNRAS.401..727S}. Stellar encounters could also
facilitate the ejection of bound brown dwarf companions from their host stars \citep{2007A&A...466..943G}. With these
additional mechanisms, disk fragmentation models would again produce more brown dwarfs in a region with higher 
stellar density, which is qualitatively what we find to be the case. However, the expected magnitude of this effect 
has not been estimated yet. 

As pointed out in Sect. \ref{s21}, the current stellar densities in IC348 and NGC1333 are probably 
representative of their primordial densities \citep{2012MNRAS.426L..11G,2012MNRAS.425..450M}. With constant
star formation rate, these should scale with the gas density in the original cloud. Under these assumptions, we 
can also put limits 
on scenarios for brown dwarf formation through turbulent fragmentation. According to the model presented by
\citet{2002ApJ...576..870P}, a factor of 5 in density enhancement should amount to a very large increase (about 
an order of magnitude, see their Fig. 1) in the number of brown dwarfs. In the gravoturbulent picture 
\citep{2009ApJ...702.1428H} the effect seems to be similar. Qualitatively the result is as seen in the
Perseus clusters (i.e. the denser cluster produces more brown dwarfs), but the magnitude of the effect is much 
larger than what we derive. However, differences in other cluster parameters, for example in the Mach number, 
could partially erase the predicted effect. Since their predictions depend heavily on initial conditions, it
is doubtful whether empirically derived IMFs can provide a meaningful test for these models.

An important caveat in our analysis is the fact that what we derive is a snapshot of the mass distribution,
which may not necessarily represent the IMF. This is particularly relevant because NGC1333 is at an earlier
evolutionary state than IC348 and might become as rich as its sibling at the other side of the Perseus star forming
complex \citep{1996AJ....111.1964L}. In the typical picture of cluster formation, however, lower mass objects
form later \citep[e.g.][their Fig. 8]{2012MNRAS.419.3115B}, thus, if additional formation processes in NGC1333
have any effect on the mass distribution, they are expected to amplify the observed discrepancy with IC348. For the
comparison with the models quoted above, which typically only predict a mass distribution of cores, not an
IMF, this issue is not of practical relevance.

\section{Summary}

We present a systematic study of the mass distribution in the two young open clusters IC348 and NGC1333,
with specific emphasis on the substellar regime. These two regions are of specific interest because NGC1333
has a higher spatial density (by a factor of 4-7). In the following we list our most important findings.

\begin{enumerate}
\item{The mass distribution as well as the parameters derived from it, e.g., the star/BD ratio $R$ or the
median mass, is significantly affected by the choice of the isochrone used to estimated masses and the choice
of the cluster parameters. Therefore, we caution against comparing IMF parameters derived using different
assumptions.}
\item{If NGC1333 is in fact closer to the Sun than IC348, as indicated by several independent studies,
there is a significant difference in the mass distributions of these two clusters, in the sense that NGC1333
harbours a larger fraction of very low mass stars and brown dwarfs.}
\item{The star/BD ratio in NGC1333 is 1.9-2.4 in NGC1333, consistent with previous estimates, and 2.9-4.0 
in IC348, significantly lower than in previous estimates. The combined uncertainty in these values is approximately
$\pm 1$, but can be lowered with more accurate distance estimates and age estimates. If confirmed, these
values would point to a larger fraction of brown dwarfs in NGC1333.}
\item{These results (2 and 3) indicate that the relative number of very low mass objects in a star forming regions
may depend on the stellar density, in the sense that regions with higher density (such as NGC1333) produce
more very low mass objects. At this point, this conclusion is only based on two clusters and needs to be verified
in other regions.}
\end{enumerate}

\acknowledgements
We thank Nickolas Moeckel, Gilles Chabrier, and Francesco Palla for helpful suggestions regarding topics 
discussed in this paper. AS acknowledges financial support through the grant 10/RFP/AST2780 from the 
Science Foundation Ireland. Additional support for this work came from grants to RJ from the Natural Sciences 
and Engineering Research Council of Canada.


\begin{thebibliography}{5}
\expandafter\ifx\csname natexlab\endcsname\relax\def\natexlab#1{#1}\fi

\bibitem[{{Allard} {et~al.}(2001){Allard}, {Hauschildt}, {Alexander},
  {Tamanai}, \& {Schweitzer}}]{2001ApJ...556..357A}
{Allard}, F., {Hauschildt}, P.~H., {Alexander}, D.~R., {Tamanai}, A., \&
  {Schweitzer}, A. 2001, \apj, 556, 357

\bibitem[{{Allard} {et~al.}(2011){Allard}, {Homeier}, \&
  {Freytag}}]{2011ASPC..448...91A}
{Allard}, F., {Homeier}, D., \& {Freytag}, B. 2011, in Astronomical Society of
  the Pacific Conference Series, Vol. 448, 16th Cambridge Workshop on Cool
  Stars, Stellar Systems, and the Sun, ed. C.~{Johns-Krull}, M.~K. {Browning},
  \& A.~A. {West}, 91

\bibitem[Alves de Oliveira et 
al.(2012)]{2012A&A...539A.151A} Alves de Oliveira, C., Moraux, E., Bouvier, J., \& Bouy, H.\ 2012, \aap, 539, A151 

\bibitem[Alves de Oliveira et 
al.(2013)]{2013A&A...549A.123A} Alves de Oliveira, C., Moraux, E., Bouvier, J., et al.\ 2013, \aap, 549, A123 

\bibitem[Andersen et al.(2008)]{2008ApJ...683L.183A} Andersen, M., Meyer, 
M.~R., Greissl, J., \& Aversa, A.\ 2008, \apjl, 683, L183 

\bibitem[Andersen et 
al.(2011)]{2011A&A...534A..10A} Andersen, M., Meyer, M.~R., Robberto, M., Bergeron, L.~E., \& Reid, N.\ 2011, \aap, 534, A10 

\bibitem[Bally et al.(2008)]{2008hsf1.book..308B} Bally, J., Walawender, 
J., Johnstone, D., Kirk, H., 
\& Goodman, A.\ 2008, Handbook of Star Forming Regions, Volume I, 308 

\bibitem[Baraffe et 
al.(2002)]{2002A&A...382..563B} Baraffe, I., Chabrier, G., Allard, F., \& Hauschildt, P.~H.\ 2002, \aap, 382, 563 

\bibitem[Bastian et 
al.(2010)]{2010ARA&A..48..339B} Bastian, N., Covey, K.~R., \& Meyer, M.~R.\ 2010, \araa, 48, 339 

\bibitem[Basu 
\& Vorobyov(2012)]{2012ApJ...750...30B} Basu, S., \& Vorobyov, E.~I.\ 2012, \apj, 750, 30 

\bibitem[Bate(2012)]{2012MNRAS.419.3115B} Bate, M.~R.\ 2012, \mnras, 419, 
3115 

\bibitem[Belikov et 
al.(2002)]{2002A&A...387..117B} Belikov, A.~N., Kharchenko, N.~V., Piskunov, 
A.~E., Schilbach, E., \& Scholz, R.-D.\ 2002, \aap, 387, 117 

\bibitem[Bonnell et al.(2008)]{2008MNRAS.389.1556B} Bonnell, I.~A., Clark, 
P., \& Bate, M.~R.\ 2008, \mnras, 389, 1556 

\bibitem[Bonnell et al.(2007)]{2007prpl.conf..149B} Bonnell, I.~A., Larson, 
R.~B., \& Zinnecker, H.\ 2007, Protostars and Planets V, 149 

\bibitem[Cambr{\'e}sy et 
al.(2006)]{2006A&A...445..999C} Cambr{\'e}sy, L., Petropoulou, V., Kontizas, M., \& Kontizas, E.\ 2006, \aap, 445, 999 

\bibitem[Cameron(2011)]{2011PASA...28..128C} Cameron, E.\ 2011, \pasa, 28, 
128 

\bibitem[{{Cardelli} {et~al.}(1989){Cardelli}, {Clayton}, \&
  {Mathis}}]{1989ApJ...345..245C}
{Cardelli}, J.~A., {Clayton}, G.~C., \& {Mathis}, J.~S. 1989, \apj, 345, 245

\bibitem[Cernis(1990)]{1990Ap&SS.166..315C} Cernis, K.\ 1990, \apss, 166, 315 

\bibitem[Chabrier(2003)]{2003PASP..115..763C} Chabrier, G.\ 2003, \pasp, 
115, 763 

\bibitem[Close et al.(2005)]{2005Natur.433..286C} Close, L.~M., Lenzen, R., 
Guirado, J.~C., et al.\ 2005, \nat, 433, 286 

\bibitem[Da Rio et al.(2012)]{2012ApJ...748...14D} Da Rio, N., Robberto, 
M., Hillenbrand, L.~A., Henning, T., \& Stassun, K.~G.\ 2012, \apj, 748, 14 

\bibitem[Dawson et al.(2013)]{2013MNRAS.429..903D} Dawson, P., Scholz, A., 
Ray, T.~P., et al.\ 2013, \mnras, 429, 903 

\bibitem[de Zeeuw et al.(1999)]{1999AJ....117..354D} de Zeeuw, P.~T., 
Hoogerwerf, R., de Bruijne, J.~H.~J., Brown, A.~G.~A., 
\& Blaauw, A.\ 1999, \aj, 117, 354 

\bibitem[Gieles et al.(2012)]{2012MNRAS.426L..11G} Gieles, M., Moeckel, N., 
\& Clarke, C.~J.\ 2012, \mnras, 426, L11 

\bibitem[Goodwin 
\& Whitworth(2007)]{2007A&A...466..943G} Goodwin, S.~P., \& Whitworth, A.\ 2007, \aap, 466, 943 

\bibitem[Gutermuth et al.(2009)]{2009ApJS..184...18G} Gutermuth, R.~A., 
Megeath, S.~T., Myers, P.~C., et al.\ 2009, \apjs, 184, 18 

\bibitem[Gutermuth et al.(2008)]{2008ApJ...674..336G} Gutermuth, R.~A., 
Myers, P.~C., Megeath, S.~T., et al.\ 2008, \apj, 674, 336 

\bibitem[Haisch et al.(2001)]{2001ApJ...553L.153H} Haisch, K.~E., Jr., 
Lada, E.~A., \& Lada, C.~J.\ 2001, \apjl, 553, L153 

\bibitem[Hirota et al.(2011)]{2011PASJ...63....1H} Hirota, T., Honma, M., 
Imai, H., et al.\ 2011, \pasj, 63, 1 

\bibitem[Helling et al.(2008)]{2008MNRAS.391.1854H} Helling, C., Ackerman, 
A., Allard, F., et al.\ 2008, \mnras, 391, 1854 

\bibitem[Hennebelle 
\& Chabrier(2009)]{2009ApJ...702.1428H} Hennebelle, P., \& Chabrier, G.\ 2009, \apj, 702, 1428 

\bibitem[Irwin et al.(2007)]{2007MNRAS.380..541I} Irwin, J., Aigrain, S., 
Hodgkin, S., et al.\ 2007, \mnras, 380, 541 

\bibitem[J{\o}rgensen et al.(2006)]{2006ApJ...645.1246J} J{\o}rgensen, 
J.~K., Harvey, P.~M., Evans, N.~J., II, et al.\ 2006, \apj, 645, 1246 

\bibitem[J{\o}rgensen et al.(2008)]{2008ApJ...683..822J} J{\o}rgensen, 
J.~K., Johnstone, D., Kirk, H., et al.\ 2008, \apj, 683, 822 

\bibitem[Kroupa(2001)]{2001MNRAS.322..231K} Kroupa, P.\ 2001, \mnras, 322, 
231 

\bibitem[Lada et al.(1996)]{1996AJ....111.1964L} Lada, C.~J., Alves, J., 
\& Lada, E.~A.\ 1996, \aj, 111, 1964 

\bibitem[Lombardi et 
al.(2010)]{2010A&A...512A..67L} Lombardi, M., Lada, C.~J., \& Alves, J.\ 2010, \aap, 512, A67 

\bibitem[Luhman(1999)]{1999ApJ...525..466L} Luhman, K.~L.\ 1999, \apj, 525, 
466 

\bibitem[Luhman et al.(2003)]{2003ApJ...593.1093L} Luhman, K.~L., Stauffer, 
J.~R., Muench, A.~A., et al.\ 2003, \apj, 593, 1093 

\bibitem[Luhman 
\& Potter(2006)]{2006ApJ...638..887L} Luhman, K.~L., \& Potter, D.\ 2006, \apj, 638, 887 

\bibitem[Moeckel et al.(2012)]{2012MNRAS.425..450M} Moeckel, N., Holland, 
C., Clarke, C.~J., \& Bonnell, I.~A.\ 2012, \mnras, 425, 450 

\bibitem[Muench et al.(2003)]{2003AJ....125.2029M} Muench, A.~A., Lada, 
E.~A., Lada, C.~J., et al.\ 2003, \aj, 125, 2029 

\bibitem[Mu{\v z}i{\'c} et al.(2012)]{2012ApJ...744..134M} Mu{\v z}i{\'c}, 
K., Scholz, A., Geers, V., Jayawardhana, R., 
\& Tamura, M.\ 2012, \apj, 744, 134 

\bibitem[Padoan 
\& Nordlund(2002)]{2002ApJ...576..870P} Padoan, P., \& Nordlund, {\AA}.\ 2002, \apj, 576, 870 

\bibitem[Prusti(2011)]{2011sca..conf...11P} Prusti, T.\ 2011, Stellar 
Clusters \& Associations: A RIA Workshop on Gaia, 11 

\bibitem[{{Schlegel} {et~al.}(1998){Schlegel}, {Finkbeiner}, \&
  {Davis}}]{1998ApJ...500..525S}
{Schlegel}, D.~J., {Finkbeiner}, D.~P., \& {Davis}, M. 1998, \apj, 500, 525

\bibitem[Scholz et al.(2009)]{2009ApJ...702..805S} Scholz, A., Geers, V., 
Jayawardhana, R., et al.\ 2009, \apj, 702, 805 

\bibitem[Scholz et al.(2012)]{2012ApJ...744....6S} Scholz, A., Muzic, K., 
Geers, V., et al.\ 2012, \apj, 744, 6 

\bibitem[Shen et al.(2010)]{2010MNRAS.401..727S} Shen, S., Wadsley, J., 
Hayfield, T., \& Ellens, N.\ 2010, \mnras, 401, 727 

\bibitem[Stamatellos \& Whitworth(2009)]{2009MNRAS.392..413S} Stamatellos, D., 
\& Whitworth, A.~P.\ 2009, \mnras, 392, 413 

\bibitem[Stassun et al.(2007)]{2007ApJ...664.1154S} Stassun, K.~G., 
Mathieu, R.~D., \& Valenti, J.~A.\ 2007, \apj, 664, 1154 

\bibitem[{{Stassun} {et~al.}(2012){Stassun}, {Kratter}, {Scholz}, \&
  {Dupuy}}]{2012ApJ...756...47S}
{Stassun}, K.~G., {Kratter}, K.~M., {Scholz}, A., \& {Dupuy}, T.~J. 2012, \apj,
  756, 47

\bibitem[Thies et al.(2010)]{2010ApJ...717..577T} Thies, I., Kroupa, P., 
Goodwin, S.~P., Stamatellos, D., \& Whitworth, A.~P.\ 2010, \apj, 717, 577 

\bibitem[Whitworth et al.(2007)]{2007prpl.conf..459W} Whitworth, A., Bate, 
M.~R., Nordlund, {\AA}., Reipurth, B., 
\& Zinnecker, H.\ 2007, Protostars and Planets V, 459 

\bibitem[Wolk et al.(2008)]{2008hsf2.book..124W} Wolk, S.~J., Bourke, 
T.~L., \& Vigil, M.\ 2008, Handbook of Star Forming Regions, Volume II, 124 

\bibitem[Wuchterl 
\& Tscharnuter(2003)]{2003A&A...398.1081W} Wuchterl, G., \& Tscharnuter, W.~M.\ 2003, \aap, 398, 1081 

\end{thebibliography}
\end{document}